\pdfoutput=1
\documentclass[acmlarge]{acmart}

\usepackage[utf8]{inputenc}
\usepackage[english]{babel} 
\usepackage{listings}
\usepackage{hyperref}
\usepackage{multirow}
\usepackage{bm}
\usepackage{amsbsy}
\usepackage{tabularx}

\usepackage{xcolor}  
\usepackage[export]{adjustbox} 
\usepackage{subfig}  
\usepackage[noframe]{showframe}  

\graphicspath{{fig/}}

\lstdefinelanguage
   [riscv]{Assembler}     
   [x86masm]{Assembler} 
   {commentstyle=\color{myDarkGreen},morekeywords={rem}} 

\definecolor{darkgrey}{rgb}{0.22,0.22,0.22}
\definecolor{myDarkGreen}{rgb}{0.0,0.5,0.0}
\definecolor{blueReview}{rgb}{0.1,0.1,0.8}

\newcommand{\dmr}[1]{%
#1}
\usepackage[
disable,
colorinlistoftodos
	]{todonotes}

\newcommand{\polen}{{\sc PolEn}}
\newcommand{\spike}{{Spike}}
\newcommand{\code}[1]{{\small\texttt{#1}}}
\newcommand{\aesttable}{\code{aes-t-table}}
\newcommand{\camellia}{\code{camellia}}
\newcommand{\des}{\code{3des}}

\newcommand{\aesb}{\code{aes-8-bits}}
\newcommand{\misty}{\code{misty}}
\newcommand{\simon}{\code{simon}}
\newcommand{\trivium}{\code{trivium}}
\newcommand{\md}{\code{md5}}
\newcommand{\sha}{\code{sha256}}

\newcommand{\plainC}{\textsc{unprotected}}
\newcommand{\encC}{\textsc{encrypted}}
\newcommand{\polyC}{\textsc{polymorphic}}
\newcommand{\polenC}{\textsc{polen}}

\newcommand{\kt}{k_\mathcal{T}}

\newcommand{\off}{\color{red}{$\boldsymbol{\times}$}}
\newcommand{\on}{\textcolor[rgb]{0,0.6,0}{\pmb{\checkmark}}}

\setcopyright{rightsretained}
\acmJournal{DTRAP}
\acmYear{2021} \acmVolume{1} \acmNumber{1} \acmArticle{1} \acmMonth{1} \acmPrice{}\acmDOI{10.1145/3487058}



\begin{document}
\title{Code Polymorphism Meets Code Encryption: 
Confidentiality and Side-Channel Protection of Software Components}

\author{Lionel Morel}
\author{Damien Couroussé}
\affiliation{%
  \institution{Univ.\ Grenoble Alpes, CEA, List}
  \city{Grenoble}
  \country{France}
  \postcode{F-38000}
}
\author{Thomas Hiscock}
\affiliation{%
  \institution{Univ.\ Grenoble Alpes, CEA, Leti}
  \city{Grenoble}
  \country{France}
  \postcode{F-38000}
}

\listoftodos{}
\newpage{}

\begin{abstract}

In this paper, we consider that, in practice, attack scenarios involving side-channel analysis combine two successive phases:
an analysis phase, targeting the extraction of information about the target and the identification of possible vulnerabilities;
and an exploitation phase, applying attack techniques on candidate vulnerabilities.
We advocate that protections need to cover these two phases in order to be effective against real-life attacks.
We present \polen, a toolchain and a processor architecture that combine countermeasures in order to provide an effective mitigation of side-channel attacks: 
as a countermeasure against the analysis phase, our approach considers the use of code encryption; 
as a countermeasure against the exploitation phase, our approach considers the use of code polymorphism, because it relies on runtime code generation, and its combination with code encryption is particularly challenging.
%
Code encryption is supported by a processor extension such that machine instructions are only decrypted inside the CPU, which effectively prevents reverse engineering or any extraction of useful information from memory dumps. 
Code polymorphism is implemented by software means.
It regularly changes the observable behaviour of the program, making it unpredictable for an attacker, hence reducing the possibility to exploit side-channel leakages. 
We present a prototype implementation, based on the RISC-V Spike simulator and a modified LLVM toolchain. 
In our experimental evaluation,
we illustrate that \polen{} effectively reduces side-channel leakages.
For the protected functions evaluated, static memory use increases by a factor of 5 to 22, corresponding to
the joint application of code encryption and code polymorphism.
The overhead, in terms of execution time, ranges between a factor of 1.8 and 4.6.

\end{abstract}

\maketitle






\section{Introduction}
\label{sec:introduction}

\dmr{Side-channel attacks (SCA) have been extensively studied in the past three decades as they are a significant threat to many computing architectures, 
but multiple challenges still lie ahead of us in order to build computing systems that are safe from such threats.
Side-channel attacks extract sensitive information from measurements of physical quantities such as power consumption or electromagnetic (EM) emanations.
The attacker relies on two, key capabilities:
i)~the observation of the physical quantities during the execution of a sensitive computation on the target; and
ii)~the analysis of measured physical quantities, often called \emph{traces}, with the aim of establishing a relationship with hypothetical sensitive values.
The side-channel research community is currently focused on the development of powerful analysis techniques and countermeasures against side-channel analysis.
}

In this paper, we focus on the practical security of software components
against side-channel attacks that are based on the observation of quantities such as power consumption or EM emanations.
Without loss of generality, we consider that an attack scenario is the combination of two, successive phases:
i)~analysis, and 
ii)~exploitation.
In the first phase, the attacker gathers information about the target to identify the elements that will be the focus of the second phase.
\dmr{Firmware extraction and its reverse engineering are the crux of the analysis phase, which has stimulated many binary analysis studies~\cite{ShoshitaishviliSOKStateArt2016}.
Once the software implementation of the target has been reverse engineered, the attacker can extend the analysis looking for software vulnerabilities~\cite{CovaStaticDetectionVulnerabilities2006}, or leverage hardware attacks to exploit software vulnerabilities~\cite{CuiBADFETDefeatingModern2017}.
Furthermore, an attacker can leverage side-channel observations to learn more about the implementation of the target:
for example, to identify which security functions are implemented, when they are launched, and on which hardware blocks they are executed~\cite{KocherIntroductiondifferentialpower2011}.
Moreover, most cryptographic primitives, if unprotected, have a specific side-channel signature that an expert can easily identify.}

\dmr{In the research literature, much attention is dedicated to the exploitation phase. 
However, it was demonstrated that reverse engineering, hence the analysis phase, is of strong importance in practical attacks (see Section~\ref{sec:background:on the importance}).
In practice, if enough implementation details of the target are known, the actual side-channel analysis is, according to \citeauthor{BronchainSideChannelCountermeasuresDissection2020}, ``\emph{usually close to trivial}''~\cite{BronchainSideChannelCountermeasuresDissection2020}.
As a consequence, practical security to protect against side-channel attacks must be supported by a first line of defence against the analysis phase. 
However, this should not be understood as an argument in favour of \emph{security by obscurity}. Instead, we assume that security evaluations in the worst-case security setting should be possible using, for example, a white-box analysis.
Ultimately, our goal is that, even if the attacker has access to the full specification of the target, they cannot easily extract meaningful information during the analysis phase.
This implies that strong cryptography is the only way to protect against a knowledgeable attacker.
In this paper, we use code encryption to ensure program confidentiality, which introduces a strong line of defence against the analysis phase.}

\dmr{Countermeasures against side-channel attacks mostly target implementations of cryptographic primitives (see Section~\ref{sec:background:SCA}).
However, many countermeasures are \emph{ad hoc}: they seek to protect one or a few primitives, and require close attention to their application.
Improving the security of a set of software components requires generic countermeasures that can be applied without specific knowledge of the components in question.
Hiding countermeasures, in particular, are interesting because unlike masking countermeasures, their application does not require dedicated knowledge of the protected component, and hence they can be used to harden various software components.
They introduce a first level of hardening against a side-channel attack, their application can be automated by the compiler~\cite{Agostacodemorphingmethodology2012,AgostaMEETApproachSecuring2015,Belleville2018}, and they can be combined with a masking countermeasure for higher levels of security~\cite{AgostaCompilerbasedTechniquesSecure2019}.}

\dmr{In this paper, we seek to identify practical and effective ways to protect embedded systems against a complete attack scenario leveraging side-channel attacks, encompassing the analysis and the exploitation phases described above.
We study a combination of code encryption (as a protection against the analysis phase) and code polymorphism (as a protection against the exploitation phase).
Code polymorphism is particularly interesting for fast and easy deployment of countermeasures as it moves the burden from the programmer to the compiler.
However, the most security-effective forms of code polymorphism involve runtime code generation.
In order to provide a full protection against the analysis phase, runtime code generation needs to support code encryption as well. 
This raises several challenges that we address in this paper.
The extension of our work to include other conventional side-channel protections, such as masking, then becomes straightforward.
}

\paragraph{Summary of contributions}
\dmr{We present \polen{}, a compiler toolchain and a processor architecture that together improve the practical security of software components in embedded systems against reverse engineering and side-channel attacks.}
\polen{} demonstrates the effective combination of two countermeasures: code encryption, and code polymorphism.

\begin{itemize}

\item
We extend the architecture of a RISC-V 32-bit processor to execute encrypted instructions, and to support runtime generation of encrypted code.
We demonstrate how runtime code generation can be combined with code encryption. 
As the runtime code generator and parts of the program that are re-generated at runtime are encrypted in memory, all forms of code extraction are prevented: offline code extraction from ROM, and online extraction at runtime.

\item 
Our support for runtime generation of encrypted code is applied to code polymorphism.
We present a full LLVM toolchain for the automatic application of code encryption and code polymorphism countermeasures.
As countermeasures are entirely implemented in the hardware, compiler and via runtime support, the burden for the developer is reduced. 

\item  
Finally, we evaluate our implementation in terms of security and performance.
In particular, the security evaluation illustrates that code encryption alone does not protect against side-channel attacks, which emphasises the importance of combining code encryption with other protections.
The cost of re-generating new polymorphic instances is also discussed. 
\end{itemize}

The rest of the paper is organised as follows.
Section~\ref{sec:background} introduces our security model, and some background knowledge concerning both code encryption and code polymorphism.
Section~\ref{sec:polen} presents our proposed combination of code polymorphism and encryption, along with an overview of \polen, which implements this combination.
Section~\ref{sec:evaluation} details our implementation, and gives an evaluation of \polen, both in terms of performance and security.
Section~\ref{sec:discussion} discusses some of the implications of our approach, and Section~\ref{sec:related works} relates it to previous works.
Finally, Section~\ref{sec:conclusion} concludes.

\section{Background}
\label{sec:background}

\subsection{Platform and Security Model}
\label{sec:platform}

In this work, we target typical IoT-grade system-on-chip (SoC) platforms comprising a microcontroller and a set of off-chip memories (e.g., DRAM and flash).
%
\dmr{The only hardware requirements of \polen{} are i)~a secure storage of encryption keys (e.g., memory, or dedicated registers) that can only be accessed by the processor;} and
ii)~a secure random number generator (RNG) that an attacker cannot probe or tamper with---this latter assumption is typical of secure systems.
%
%
%
We assume that an attacker has the following capabilities:
they can read the content of off-chip memories (DRAM and flash), and dump program code stored in these memories; and
%
\dmr{they can perform side-channel attacks, e.g., by probing the chip for EM emanations or power consumption. }
\dmr{Timing attacks and logical side-channels (cache attacks, branch prediction, ...) are out of the scope of this work.}
Finally, we assume no particular protection of data in memory.
This is an important aspect of a system's security in general, which is left out of the scope of this work.

\subsection{\dmr{The Analysis Phase and Protections Against Reverse Engineering}}

Reverse engineering encompasses a variety of techniques that can be used to retrieve a meaningful source code representation of a program that is otherwise only accessible in binary form.
An attacker usually starts by \dmr{disassembling} the extracted binary code to obtain an assembly version.
Then, they try to decompile it, in order to build a high-level source code version of the intended behaviour. 
%
Various techniques have been proposed to counter reverse engineering.
Obfuscation is a set of source code modification techniques that aim to make the behaviour of a program unintelligible~\cite{Collberg2009,Barak2016}. 
Its main advantage is that it is a software-only technique and, thus, is easy to apply to an existing code base and port to a wide range of targets. 
%
Many software approaches have been proposed to counter reverse engineering, among which Instruction Set Randomisation (ISR) has been widely studied~\cite{Barrantes2003,Cheng2019}. 
The idea underlying the latter technique is to dynamically change opcode encodings, making the instruction memory harder to decode.
These techniques can be applied by a virtual machine, an interpreter or a processor.
Unfortunately, these protections do not resist real-life code injection or code-reuse attacks~\cite{Shacham2007,Pappas2012,Sinha2019}.

Memory encryption~\cite{Henson2014} is arguably the strongest countermeasure against reverse engineering. 
Security is founded on robust cryptographic constructions: even if an adversary can access memory, they would still have to know the secret key to access data.
Memory encryption can be implemented entirely in software, demonstrated by the full-disk encryption that is provided by common operating systems.
Usually, data is decrypted while being transferred from persistent to dynamic memory (e.g., flash to DRAM).
A stronger option is to encrypt data until it reaches the memory ports of the CPU~\cite{Suh2005},
although it should be noted that this requires an in-depth re-design of the memory architecture, which is not easy with complex architectures.

Code encryption hardens programs against reverse engineering, as well as code reuse and code injection attacks.
Here, programs are encrypted before deployment and remain encrypted in memory.
Code encryption can be performed at the granularity of a memory page.
\citeauthor{Sinha2017} propose extensions to the memory architecture that are specifically adapted to both the architecture (in particular the memory management unit) and the OS (which manages encryption keys)~\cite{Sinha2017}.
However, this approach assumes the use of virtual memory, making it difficult to port to IoT-grade microcontrollers.
Code decryption can be performed on an instruction-by-instruction basis, within the CPU, after an instruction is fetched~\cite{Werner2018,Sinha2017,Hiscock2019}.
All such approaches require modifications to the CPU's micro-architecture, and thus need to be anticipated in the processor design flow.
We adopt this latter approach to code encryption in \polen{}:
%
our encryption scheme is based on the work of \citeauthor{Hiscock2019}~\cite{Hiscock2019}, which strikes a balance between portability, including to low-grade IoT-like platforms, and the ability to trade-off security and performance, by manipulating programs from within the compiler.
\polen{} can also be adapted to support the encryption scheme used by \citeauthor{Werner2018}, which supports code confidentiality and control-flow integrity~\cite{Werner2018}.


\subsection{\dmr{On the Importance of the Analysis Phase in a Side-Channel Attack}}
\label{sec:background:on the importance}

\dmr{Worst-case security models assume that the attacker has detailed knowledge of the target, which lessens the importance of the analysis phase.
In certification schemes, evaluators are given detailed knowledge of the security target, and verify the compliance of its description~\cite{ANSSICommonCriteriaCertification2017}.
Similarly, most research assumes full knowledge, and control over the evaluated target.
As a consequence, far less attention is paid to mitigation against the analysis phase than the exploitation phase.
However, several articles that describe the use of side-channel analysis in practical attack scenarios underline the importance of reverse engineering.
For example, \citeauthor{BronchainSideChannelCountermeasuresDissection2020} analyse an implementation of AES hardened against side-channel attacks~\cite{BronchainSideChannelCountermeasuresDissection2020}. 
After a preliminary in-depth investigation of the implementation of the countermeasures, 
they identify an efficient side-channel analysis to carry out in the exploitation phase.
In a similar vein, \citeauthor{OswaldWhenReverseEngineeringMeets2013} extract the secret key of a digital locking system using an EM side-channel attack~\cite{OswaldWhenReverseEngineeringMeets2013}.
Most effort was required to reverse engineer the hardware, then the software implementation of the firmware. 
The extraction and detailed analysis of the firmware is described by the authors as an essential step before they could exploit side-channel analyses.
Interestingly, they report that the exploitation phase of the attack was able to extract the secret key from the product based on a small (150) number of observations. 
This again underlines that the exploitation phase of a side-channel attack is often low cost, once the implementation details of the target are known.
Recently, \citeauthor{LomneSideJourneyTitan2021} reported a vulnerability in the Google Titan Security Key's secure element, based on a side-channel analysis, supported by power and EM measurements~\cite{LomneSideJourneyTitan2021}.
Although the exploitation of the vulnerability requires some expertise in cryptanalysis, their paper again underlines the amount of efforts spent in reverse engineering the target. }

\subsection{\dmr{The Exploitation Phase and Protections Against Side-Channel Attacks}}
\label{sec:background:SCA}
\dmr{SCA techniques exploit the observable behaviour of the attacked program.
The attacker can use any physical quantity that can be measured on the target system: EM emanations, power consumption or even sound.}
Typically, measurements, often called \emph{traces}, are analysed to recover secret keys used in cryptographic computations.
\polen{} protects sensitive data (typically encryption keys) from SCAs.

The efficiency of side-channel analysis relies on the ability of the attacker to gather multiple traces from the target.
They then try to correlate these traces with hypothetical intermediate values of the target program (e.g., AES encryption).
The value that correlates best with observations usually corresponds to the secret key.
If there are several plausible key values, the attacker can proceed to an exhaustive enumeration to identify the secret key.

Protections against SCA  
usually fall into two categories: \emph{masking} or \emph{hiding}~\cite{MangardPoweranalysisattacks2007}.
Masking splits sensitive variables used in a computation into multiple shares, and randomisation is used to make each share statistically independent from the others.
Hiding techniques will either try to reduce leakage or add noise to it to make the attack significantly harder.
%
%
Many countermeasures that are based on software hiding involve some form of execution diversity in order to produce side-channel observations that are difficult to correlate to the secret data.
Some approaches statically pre-compute different versions of the same function~\cite{AgostaMEETApproachSecuring2015,VanCleemput2017}.
At runtime, every time the protected function is executed, one of these versions is randomly selected.
Another approach consists in inserting so-called \emph{chaff} instructions in between normal instructions~\cite{Agosta2018}.
The aim is to introduce fake key values that could exhibit higher correlation values than the secret key.
%
%

\emph{Code polymorphism} is probably the most powerful form of software hiding~\cite{Agostacodemorphingmethodology2012,CourousseRuntimeCodePolymorphism2016}.
The core idea is to regularly generate new versions of secure code, called \emph{polymorphic instances}, by means of runtime code generation driven by random data.
All of these polymorphic instances are functionally equivalent, but differ in their implementation, such that each execution leads to a different (side-channel) observation.
Used alone, code polymorphism raises the bar for a side-channel attacker. 
Moreover, the behavioural variability it provides could make it more difficult to reverse engineer the protected component in the analysis phase of an attack.
However, further discussion of this point is beyond the scope of our paper.
Code polymorphism as used in \polen{},  \dmr{i.e., as a countermeasure against side-channel attacks}, has been discussed and evaluated in previous work~\cite{Belleville2018}.
In the present paper, we focus on the combination of code encryption and runtime code generation, which supports code polymorphism.

  


\section{\polen}
\label{sec:polen}

\subsection{General Approach}
\label{sec:approach}

\polen{} is composed of two main components:
a dedicated compiler toolchain that supports runtime code generation (it also includes a minimal execution runtime), and an extended processor architecture that provides hardware support for the execution of encrypted code.
Code encryption and code polymorphism are automatically applied by the toolchain. These countermeasures can be applied to any source-level target function, either independently or in combination. The user can select target functions to harden at compile time, thanks to dedicated compiler options or source code annotations.
Code polymorphism is implemented for each target function by a runtime machine code generator, referred to as a Specialised Generator of Polymorphic Code (SGPC), which produces multiple, new polymorphic instances of code that implement the desired functionality (Section~\ref{sec:code polymorphism}).
%
Code encryption is implemented at compile time by the toolchain and a
post-compilation patching tool (Section~\ref{sec:encryption}).
When combined with code polymorphism, code encryption requires dedicated runtime support (Section~\ref{ssec: encrypted poly code}).

In the remainder of this paper, we use the term \textit{static} to refer to compile-time code transformations, while the term \textit{dynamic} refers to the \textit{runtime} technique used to generate polymorphic instances.

\begin{figure}
  \centering
  \includegraphics[scale=0.8]{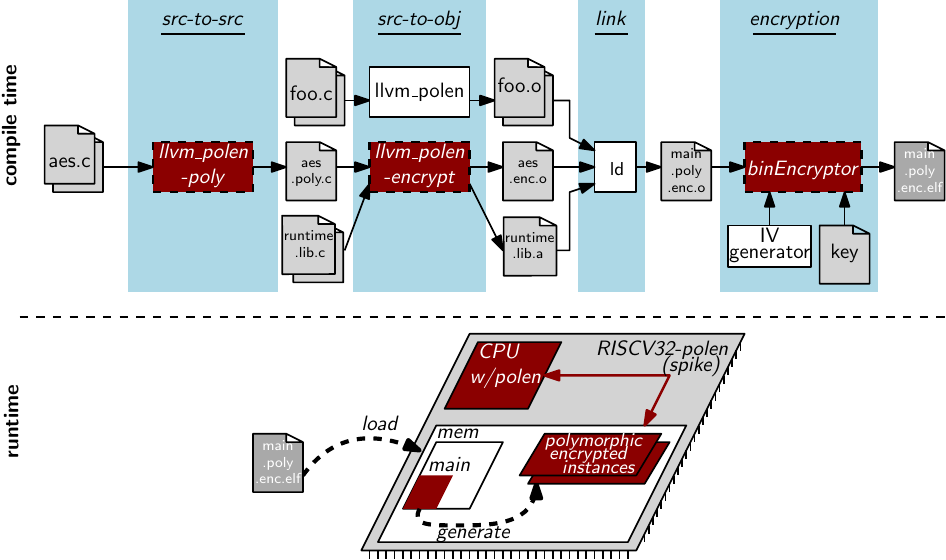}
  \caption{Overview of the \polen{} approach: complete toolchain and runtime execution.  The input file {\small \code{aes.c}} represents source code to be hardened by \polen{}, and {\small \code{foo.c}} is other application code with no security requirements.}
  \label{fig:complete toolchain detail}
\end{figure}

The \polen{} architecture is shown in Figure~\ref{fig:complete toolchain detail}.
We begin with a coarse-grained overview of the framework, the implementation details are presented in Section~\ref{sec:setup}.
Code hardening starts with a set of functions that are identified as critical for the system's security.
In the example, these are contained in the {\code{aes.c}} file.
The architecture produces a binary code \code{main.poly.enc.elf} with the desired protection.
The compile time process is divided into four phases.
In the first phase (\code{src-to-src}), the compiler, \code{llvm\_polen}, which is called with the option \code{-poly}, generates an SGPC of each function to be hardened by code polymorphism.
In the second (\code{src-to-obj}), the compiler generates an object file for each input file, and prepares the target functions for encryption.
If necessary, code encryption is also applied to parts of the runtime library (\code{runtime.lib.c}). The third phase (\code{link}) is an unmodified link stage that uses the standard GNU linker. Finally, code encryption is performed at the binary level by a separate tool (\code{binEncryptor}), which produces encrypted code using device-specific keys. 
\polen{} is highly configurable.
The user can choose to apply either code encryption or code polymorphism, or combine them at different granularities.
For example, it can be used to encrypt polymorphic instances, but not the SGPC itself.

The runtime part of \polen{} is a combination of hardware and software components.
On the software side, the SGPCs periodically produce new polymorphic instances in the form of binary code.
If both encryption and code polymorphism are applied, these instances are encrypted.
In the latter case, encryption is handled by the SGPC, before code is written to program memory and executed.
On the hardware side, the CPU is modified to include support on-the-fly decryption of instructions.
This allows encrypted programs to be executed, both those generated statically (e.g., the SGPC or other procedures without code polymorphism) and dynamically (polymorphic instances).
The hardware is extended with encryption support for the code emission of encrypted polymorphic instances as they are produced by the SGPC (Section~\ref{sec:encrypted polymorphic code:hardware support}).


\subsection{Code Polymorphism}
\label{sec:code polymorphism}

When code polymorphism is applied to a target function at the source level, the original function is replaced by a \emph{wrapper} and an SGPC (\code{src-to-src}, Figure~\ref{fig:complete toolchain detail}).
The wrapper has the same prototype as the original target function. It encapsulates the SGPC such that a call to a polymorphic function is transparent from the point of view of the rest of the program. At runtime, the wrapper i)~executes the SGPC to generate a new polymorphic instance, according to a regeneration policy defined by the user; and ii)~executes the polymorphic instance. These two phases can be re-ordered interchangeably.

Therefore, a call to the wrapper function always leads to the execution of code that has the same functionality as the original function.
Each generated instance is functionally equivalent to the original function, but uses different code shapes in order to confuse the attacker. Several code transformations can be produced by the SGPC. 

The SGPC is a C program that is specific to the current target polymorphic function. 
Its only input is random data (typically, from a RNG) that is assumed to be out of the control of an attacker.
The SGPC generation is managed by a dedicated compiler backend that emits code written in C rather than assembly or object code.
The original machine instructions are replaced by calls to instruction-level code generation functions that handle variability.

Runtime variability is handled as follows:
at the instruction level, the SGPC can introduce a sequence of instructions that are a \textit{semantically equivalent variant} of the original instruction.
%
At the basic-block level, \textit{instruction shuffling} can re-order independent instructions.
At the function level, \textit{register shuffling} can be performed. 
Each new generation produces a new permutation of registers, and this new permutation is used to generate all of the instructions in the current instance. 
At the basic-block level, \textit{noise} can be inserted via carefully-chosen instructions that do not alter the program's behaviour. A register liveness analysis is performed that allows noise and functional instructions to share the same set of registers. The amount of noise to be inserted is determined each time a new instance is generated, and is determined separately for each location in the code where noise is inserted.
We do not provide more details about code polymorphism here, as the approach has already been described in full, and evaluated in the literature~\cite{Belleville2018}.

\begin{figure}
  \centering
  \includegraphics[width=0.9\textwidth]{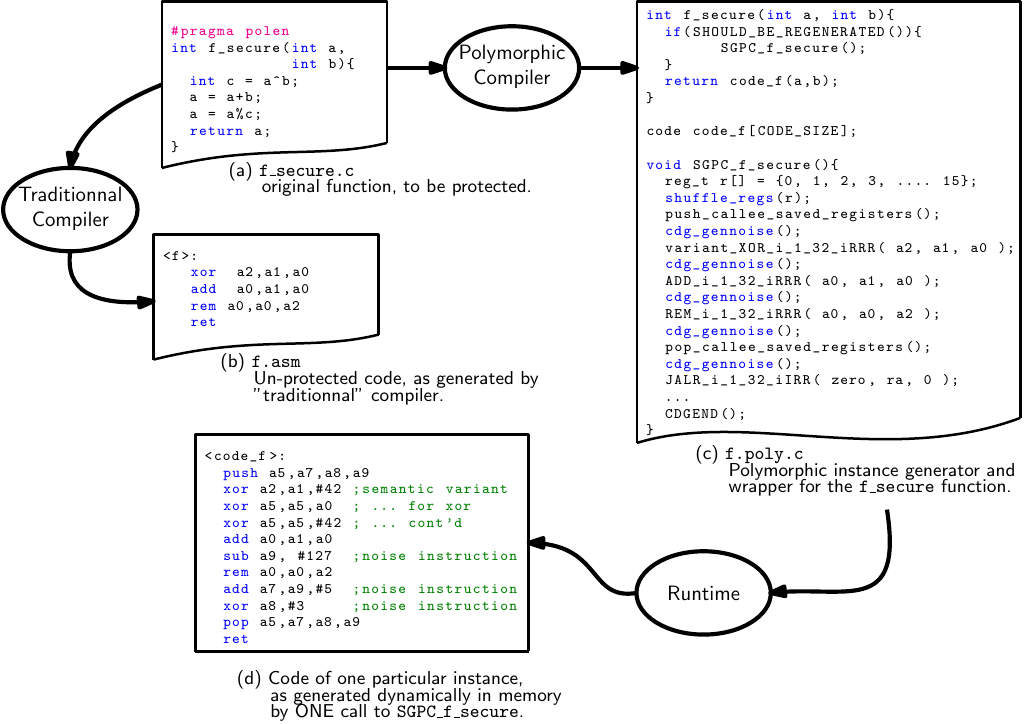}
  \caption{Code polymorphism: general overview. }
  \label{fig:example code cogito}
\end{figure}

\paragraph{Example}

Figure~\ref{fig:example code cogito} shows the process for an example function \code{f\_secure}.
In Figure~\ref{fig:example code cogito}(a), the programmer identifies the function \code{f\_secure} to be protected with the \code{polen}~pragma.
Figure~\ref{fig:example code cogito}(b) shows the unprotected code that the compiler would normally generate for \code{f\_secure}. 
The \polen{} toolchain statically generates the code \code{f.poly.c} shown in Figure~\ref{fig:example code cogito}(c).
The original \code{f\_secure} function is replaced by a \textit{wrapper}, \code{f\_secure}, that is called in place of the original function.
Every time it is called, this wrapper checks if a new polymorphic instance should be generated. 
Runtime code generation is performed by the \code{SGPC\_f\_secure} function, which emits polymorphic instances in the \code{code\_f} buffer.
The wrapper then jumps to the \code{code\_f} code and executes it. 
Figure~\ref{fig:example code cogito}(d) shows one instance of code that is generated dynamically. 

The structure of the function \code{SGPC\_f\_secure} follows the structure of the function \code{f\_secure}.
For each instruction in the expected assembler code, there is a corresponding call to an instruction-level generator.
For example, the instruction \code{add a0, a1, a0} corresponds to the macro \code{ADD\_i\_1\_32\_iRRR(a0,a1,a0)}.
For some instructions, including \code{add}, these encoding macros produce the binary of the corresponding instruction.
For others, macros generate semantic variants: a call to \code{variant\_XOR\_i\_1\_32\_iRRR(a2, a1, a0)} may introduce one of many variants, instead of the original instruction.

\code{SGPC\_f\_secure} also includes calls to \code{cdg\_gennoise()}, \code{shuffle\_regs()} and \code{CDGEND()} functions from the \polen{} runtime API. Respectively, these generate noise instructions, perform register shuffling and carry out accounting when polymorphic instances have been generated (Section~\ref{ssec: encrypted poly code}).

\subsection{Code Encryption}
\label{sec:encryption}

Code encryption is a countermeasure that is implemented to protect against memory extraction and reverse engineering.
The idea is to encrypt code (off-line) using a symmetric cryptographic primitive and keep the secret key (required for decryption) inside the CPU. Thus, an attacker who succeeds in extracting the contents of memory still needs to break the cryptographic primitive in order to recover any useful data.

\subsubsection{Formalism}\label{sec:encryption_formalism}


For the sake of generality, we introduce Definition~\ref{def:encryption_framework} as an abstract representation of an encryption primitive. We describe \polen{} with respect to this formalism, in order that the description remains independent of the underlying primitive.


\begin{definition}\label{def:encryption_framework}
We define an instruction-level encryption primitive as the 3-tuple of keyed functions $(\mathcal{I}, \mathcal{T}_{enc}, \mathcal{T}_{dec})$:
\begin{itemize}
    \item $\mathcal{I}_k(IV) \mapsto state$ is the cipher \emph{initialisation function}. From a key $k$ and an initialisation vector ($IV$), it generates an initial state.
    \item $\mathcal{T}_{enc, \: k}(state, m) \mapsto (state', c)$ is the \emph{encryption function}. From the cipher current state and an input message, it produces a new state and a ciphertext.
    \item $\mathcal{T}_{dec, \: k}(state, c) \mapsto (state', m)$ is the \emph{decryption function}. From the cipher current state and an input ciphertext, it produces a new state and a plaintext.
    \item The secret key $k$ has a width of $\lambda$ bits.
    \item $state$ is a fixed-length vector that allows the encryption of arbitrary-length sequences. 
\end{itemize}
\end{definition}

The functions $(\mathcal{I}, \mathcal{T}_{enc}, \mathcal{T}_{dec})$ from Definition~\ref{def:encryption_framework} operate on fixed-length input messages (e.g., 128 bits). In order to support variable-length messages, an input message is divided into chunks of fixed length and processed block-by-block by chaining the cipher state.

In this section, we assume that messages are aligned with the chunk size.
Section~\ref{sec:block ciphers} discusses such assumptions.
The example below shows how a message made of $n + 1$ parts would be encrypted and decrypted:
\begin{center}
    \begin{tabular}{p{6cm}p{6cm}}
        Encryption & Decryption\\
        \hline
        $\mathcal{I}_k(IV) \to state_0$ & $\mathcal{I}_k(IV) \to state_0$\\
        $\mathcal{T}_{enc, \: k}(state_0, m_0) \to (state_1, \textcolor{red}{c_0})$ & $\mathcal{T}_{dec, \: k}(state_0, \textcolor{red}{c_0}) \to (state_1, m_0)$\\
        $\mathcal{T}_{enc, \: k}(state_1, m_1) \to (state_2, \textcolor{red}{c_1})$ & $\mathcal{T}_{dec, \: k}(state_1, \textcolor{red}{c_1}) \to (state_2, m_1)$\\
        $\vdots$ & $\vdots$ \\
        $\mathcal{T}_{enc, \: k}(state_n, m_n) \to (state_{n + 1}, \textcolor{red}{c_n})$ & $\mathcal{T}_{dec, \: k}(state_n, \textcolor{red}{c_n}) \to (state_{n + 1}, m_n)$\\
    \end{tabular}
\end{center}

In the context of \polen{}, we require the additional property that an encrypted instruction can be patched after code emission.
Formally, we assume that there exists a pair of functions ($\oplus$, $\odot$) such that:
\begin{equation}\label{eq:homomorphic_like}
    \mathcal{T}_{enc, \: k}(state, m_{0} \oplus c) = \mathcal{T}_{enc, \: k}(state, m_{0}) \odot c
\end{equation}
Any primitive that encrypts and decrypts using an XOR with a pseudo-random sequence (such as stream ciphers or AES in counter mode) satisfies this requirement, where the functions $\oplus$ and $\odot$ are both the binary {\tt xor} function.

An on-the-fly decryption of a list of encrypted instructions requires the cipher to be in the correct state.
In the formalism of Definition~\ref{def:encryption_framework}, there are two ways to generate a valid state:
\begin{enumerate}
    \item start a new sequence from an initialisation vector $IV$ by calling the $\mathcal{I}_k(IV)$ function, or
    \item extend an existing sequence by calling $\mathcal{T}_{enc}(state, m)$ (or $\mathcal{T}_{dec}$), which updates the cipher state.
\end{enumerate}

When the processor executes sequential instructions, extending the sequence (i.e., calling $\mathcal{T}_{dec}$) is the default choice.
\dmr{However, when a control-flow instruction is taken (be it a conditional or unconditional branch, a jump, call, or ret instruction), the correct state (i.e., the one used during encryption) must be restored according to the current execution path to decrypt the target instruction.}
This observation suggests that encryption should be applied at the granularity of basic blocks in the source program.
\dmr{We proceed as follows: an $IV$ is associated with each basic block and encryption starts with the state $\mathcal{I}_k(IV)$.}
We choose to store IVs at the beginning of every basic block so that decryption can generate the initial state.
\dmr{For each control-flow instruction taken, the CPU reads the IV at the target destination, and triggers a reset of the cipher state by calling the $\mathcal{I}_k(IV)$ function for the destination basic block.}
%


%
%
%

\subsubsection{\dmr{Control-Flow Graph (CFG) Preparation}}
\label{ssubsec:cfg prep}

\begin{figure}
  \centering
  \includegraphics[width=\textwidth]{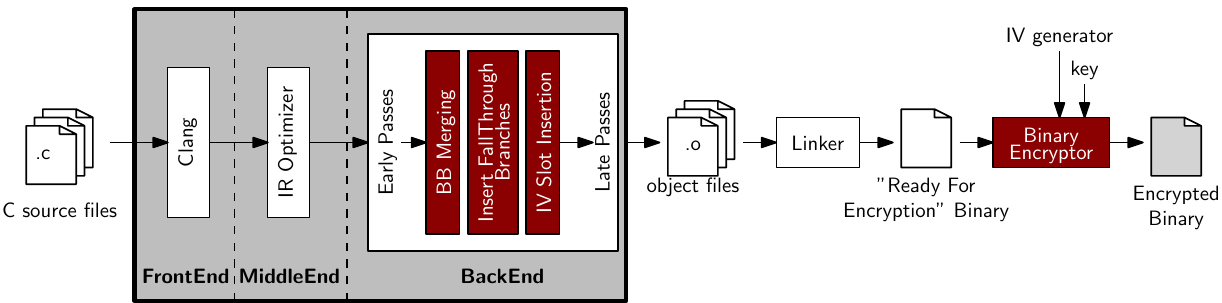}  
  \caption{The program encryption workflow. Elements related to \polen{} are shown in red boxes.}
  \label{fig:encryptionFlow}
\end{figure}

Here we describe the modifications to the target program, required to support our code encryption scheme. These modifications are performed at compile time before the application of encryption by the \code{binEncryptor} tool.
They are implemented as a series of passes that are added to the compiler's backend. The overall compilation flow is shown in Figure~\ref{fig:encryptionFlow}.

\begin{figure}
  \centering
  \subfloat[Initial BB Fallthrough\label{fig:bbBeforeMerging}]{
    \quad\includegraphics[width=.15\textwidth,keepaspectratio,valign=t]{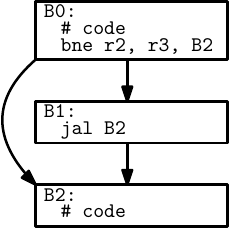}\quad
  }
  \hfill
  \subfloat[b][After Merging\label{fig:bbAfterMerging}]{
    \quad\includegraphics[width=.15\textwidth,keepaspectratio,valign=t]{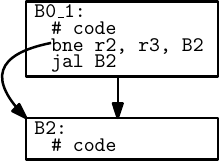}\quad
  }
  \hfill
  \subfloat[b][Initial BB Fallthrough\label{fig:fallThrough}]{
    \quad\includegraphics[width=.15\textwidth,keepaspectratio,valign=t]{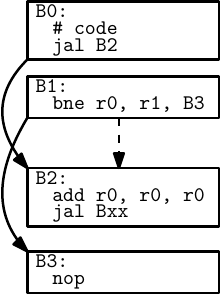}\quad
  }
  \hfill
  \subfloat[b][Corrected BB Fallthrough\label{fig:fallThrough_fixed}]{
    \quad\quad\quad\includegraphics[width=.15\textwidth,keepaspectratio,valign=t]{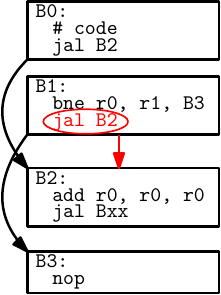}\quad\quad\quad
  }
  \caption{Example of modifications of the CFG, required for efficient encryption~\cite{Hiscock2019}.}
  \label{fig:bb transformations}
\end{figure}

First, we reduce the number of basic blocks by applying the following merging strategy.
Consider two basic blocks B1 and B2 such that B1 jumps to B2 (see Figure~\ref{fig:bbBeforeMerging} and Figure~\ref{fig:bbAfterMerging}). 
B1 and B2 can be encrypted/decrypted by the same encryption sequence if B1 is the only predecessor of B2.
In this case, our compiler merges B1 and B2.
This pass is optional, but enabling it can bring significant performance improvements if the initialisation function $\mathcal{I}_k$ of the cipher significantly increases execution time.



\dmr{A second pass ensures that all of the basic blocks with multiple predecessors are only reachable via an explicit control-flow instruction.}
\dmr{Earlier compiler passes can remove unnecessary control-flow instructions, especially when a basic block B2 is the fallthrough of a predecessor B1, as illustrated in Figure~\ref{fig:fallThrough}. As execution at the end of B1 reaches its successor B2 without a control-flow instruction, the cipher can take different state values when reaching B2.}
\dmr{Thus, if needed, the second pass appends a direct control-flow instruction to each basic block (see Figure~\ref{fig:fallThrough_fixed}). These additional control-flow instructions instruct the CPU to reset the cipher state to ensure the correct decryption of instructions.}

%

A third pass adds a slot for an IV at the start of each basic block.
The upper parts of the IV slot are filled with a special value that does not map to an instruction, and which can be matched by the \code{binEncryptor} program (Section~\ref{ssubsec:program encryption}).
The rest of the IV is filled with the number $nb_I$ of instructions contained in the basic block.
The whole IV slot is replaced by a full IV during binary encryption.

%
%

\subsubsection{Encryption}
\label{ssubsec:program encryption}

The binary produced by the compiler is then fed to a binary encryption tool, along with an encryption key (see far right of Figure~\ref{fig:encryptionFlow}).
The encryption tool scans the binary file for IV slots.
Whenever it reaches a slot: 
i)~it reads the last part of the slot as the number of instructions $nb_I$ contained in the upcoming block;
ii)~it generates a new random IV value;
iii)~it fills the current IV slot with it;
iv)~it initialises its own software implementation of the cipher, by calling $\mathcal{I}_k$ with this IV; and
v)~it encrypts the $nb_I$ instructions following the IV slot. 
%

\subsubsection{Execution of Encrypted Code}
\label{subsec: dec enc hw support}

\begin{figure}
  \centering
  \subfloat[Modifications to the processor's pipeline \textsc{Fetch} stage to decrypt encrypted instructions. \label{fig:decoding trivium}]{
    \includegraphics[scale=0.65,valign=b]{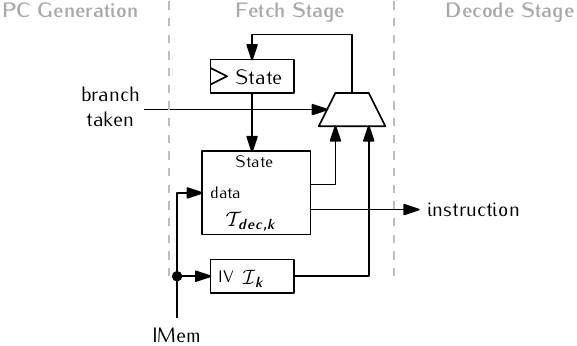}
  }
  \qquad
  \subfloat[Modifications to the processor's \textsc{Execute} stage to encrypt instructions. \label{fig:encoding trivium}]{
    \includegraphics[scale=0.65,valign=b]{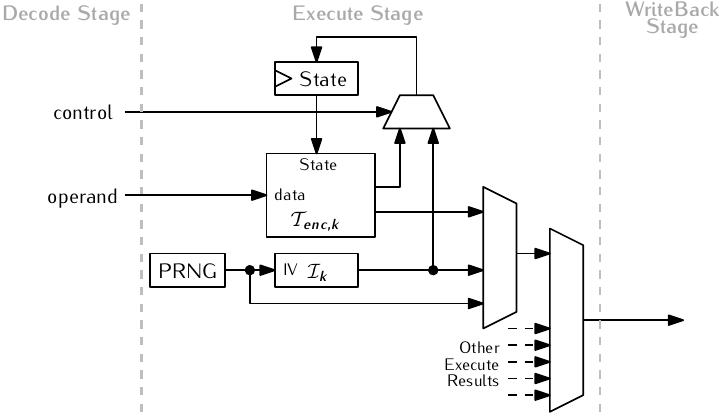}
  }
  \caption{\polen{} modified hardware.}
  \label{fig:modified hardware}
\end{figure}

Encrypted instructions are decrypted on the fly, as they are fetched by the CPU.
The processor is modified as shown in Figure~\ref{fig:decoding trivium}.
%
%
The \textsc{Fetch} stage incorporates a decryption module, labelled $\mathcal{T}_{dec, \: k}$. 
By default,
$\mathcal{T}_{dec,\: k}$ produces the plain-text instruction from the encrypted instruction obtained from memory.
The result is fed to the processor's \textsc{Decode} stage.
$\mathcal{T}_{dec, \: k}$ also generates a new cipher state from its input, and is now ready to decrypt the next instruction.
\dmr{If a control-flow instruction is taken, 
 the CPU first checks if an encryption disable signal is active, and switches to plain-text execution mode, if needed. 
If decryption continues, the cipher state is initialised by running the $\mathcal{I}_k$ module with an IV read in the program's memory.
The execution then continues normally by calling $\mathcal{T}_{dec,\: k}$.}

\subsubsection{Calling Encrypted Functions}

\dmr{A call/return to/from two encrypted functions works exactly like other control-flow instructions.}
\dmr{The callee's code starts with an IV and the corresponding control-flow instruction from the caller resets the cipher state with this IV.}
Returning from the callee, the cipher state needs to be reset with a correct state.
An IV is inserted by the compiler after each call.
%

\subsubsection{Calling Unencrypted Functions}
\label{subsec: encryption domain}


The performance/security compromise can be fine-tuned by the programmer. 
They can define functions that can be called from secured code, but that do not need to be protected themselves.
To allow this to happen, the instruction set architecture (ISA) is extended with two instructions \code{disable\_enc} and \code{enable\_enc} which, respectively, disable and enable encryption. 
In practice, the programmer only needs to specify functions that need protection.
The compiler then generates specific call/return sequences for unprotected functions, following the approach described in the following. 

\begin{figure}
  \includegraphics{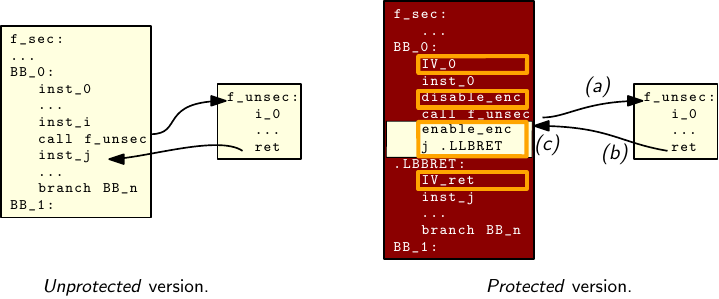}
  \caption{Handling calls to unprotected functions. When calling an unprotected function \texttt{f\_unsec} from a secured function \texttt{f\_sec}, encryption needs to be disabled and re-enabled from the calling context. Encrypted code is shown in dark red and non-encrypted code is shown in light yellow.} 
\label{fig:protected call}
\end{figure}

%
Consider the example given in Figure~\ref{fig:protected call}.
The dark red (resp.\ light yellow) block indicates that the corresponding code is encrypted (resp. not encrypted).
Instructions and IVs added to the original code are highlighted by the orange rectangle.

The secured function \code{f\_sec} calls an unprotected function \code{f\_unsec}.
The left side of Figure~\ref{fig:protected call} shows the initial state.
Here, the call to \code{f\_unsec} is simply a matter of jumping to the first instruction of the function.

When \code{f\_sec} is encrypted and \code{f\_unsec} is unencrypted, special care is taken to prepare \code{f\_sec} (see the right-hand side of Figure~\ref{fig:protected call}).
Prior to the call to \code{f\_unsec}, the processor is decrypting \code{f\_sec}'s code. 
\dmr{Before executing the call to \code{f\_unsec}, the processor is set to switch to non-decrypting on the next control-flow instruction. }
\dmr{This change of state is triggered by the \code{disable\_dec} instruction before the call.}
The call itself \textit{(a)} effectively switches the processor to its non-decrypting state, and execution of \code{f\_unsec} then continues, without decryption.
When returning from \code{f\_unsec} \textit{(b)}, the processor returns to its decrypting state.
This action cannot be triggered from \code{f\_unsec}, since the compiler does not necessarily know about its calling context. 
Thus, \code{f\_sec} resets the processor to the correct decrypting state.
\dmr{This reset is performed by inserting the sequence of two instructions labelled \textit{(c)}.}
\code{enable\_dec} switches back to the decryption state during the added jump.
\dmr{This control-flow instruction switches the processor to its decrypting state, and begins execution of an extra basic block that starts with \code{inst\_j}, the instruction that follows the call to \code{f\_unsec} in the original code.}
Together, three instructions are added to the call sequence: \code{disable\_enc}, \code{enable\_enc} and a jump to the added basic block.

The question of determining which functions should be encrypted or not is discussed in Section~\ref{sec:discussion enc scope}. 

\subsection{Encrypted Polymorphic Code}
\label{ssec: encrypted poly code}

We now describe in detail how code polymorphism and code encryption are combined in \polen. 
\polen{} is highly configurable, and the programmer can choose to activate each countermeasure \dmr{independently on parts of the protected program.
E.g., one can selectively} encrypt parts of the code.
This is discussed in Section~\ref{sec:configurability}.
In the following, we present the full combination of code polymorphism and encryption.

\subsubsection{Static Code Generation}

By default, all static code produced by the compiler for secured functions is encrypted.
For polymorphic code, this includes wrappers, SGPCs and the entire \polen{} runtime library. The latter is prepared for encryption separately, then linked with each application.
The whole binary can then be encrypted using the application's key.

\subsubsection{Dynamic Generation of Encrypted Code}

Any code produced dynamically, i.e.\ polymorphic instances, needs to be encrypted on the fly.
This is particularly important as an attacker with read access to memory could easily dump it at a well-chosen time when an instance of the \code{f\_secure} function has just been produced, and reverse engineer it.
This process is implemented as a modification of the polymorphic code generation technique described in Section~\ref{sec:code polymorphism}. Instead of the polymorphic instance shown in Figure~\ref{fig:example code cogito}, it produces an encrypted polymorphic instance, as shown in Figure~\ref{fig:enc poly instance}.
At the beginning of each basic block (including the beginning of the function itself), a new IV is inserted.
Then, each instruction generated by the SGPC is encrypted on the fly. 
Figure~\ref{fig:enc poly instance} shows, for each instruction, the sequence of encrypted words and a comment indicating the instruction it corresponds to. 

We now describe how these encrypted instances are generated, together with the required hardware and software support included in \polen{}. 

\subsubsection{Hardware Support}
\label{sec:encrypted polymorphic code:hardware support}

From the hardware point of view, we extend the \textsc{Execute} stage of our processor with two modules:
an encryption module, denoted $\mathcal{T}_{enc,\: k}$; and a cipher state initialisation module, denoted $\mathcal{I}_k$. Both are shown in Figure~\ref{fig:encoding trivium}. A pseudorandom number generator block is used to generate fresh IVs, which are used to generate an initial cipher state through $\mathcal{I}_{k}$ and are written to memory at the beginning of new basic blocks.
Decrypted instructions obtained from the decode stage (e.g., read from registers) are passed to $\mathcal{T}_{enc, \: k}$ to obtain the encrypted version.

This hardware support is accessed from software through an ISA extension that includes four new instructions, which are described in Table~\ref{tab:ISA extension}. This extension includes instructions that are used specifically to generate polymorphic instances, and others that are needed to activate/de-activate encryption, as presented in Section~\ref{subsec: encryption domain}.

\begin{table}
  \caption{ISA extension for \polen. }
  \label{tab:ISA extension}
  \centering
  \begin{tabularx}{\textwidth}{|c|X|}\hline
  \code{initBB} & \dmr{This instruction is called when emitting the code of a new basic block, it: i)~initialises the cipher state with a new, randomly-chosen IV; and ii)~writes this IV at the beginning of the generated basic block.} \\\hline
 
  \code{enc\_word} & \dmr{This instruction encrypts the content of a register and writes the corresponding encrypted word into another register. It is used to encrypt instructions of polymorphic instances.}  \\\hline
  \code{enable\_dec} & \dmr{This instruction enables decryption of incoming instructions.  Decryption starts when the following control-flow instruction is executed. The target of the control-flow instruction is, thus, expected to be encrypted.} \\\hline
  
  \code{disable\_dec} & \dmr{This instruction disables decryption of incoming instructions. As \code{enable\_dec}, decryption is disabled when the following control-flow instruction is executed.  The target of the control-flow instruction is expected not to be encrypted.}\\\hline
  \end{tabularx}
\end{table}

\subsubsection{Software Support}
\label{sec:encrypted instruction generation}

From the software point of view, the generation of encrypted polymorphic instances is implemented within the \polen{} runtime library.
Encryption is performed on a per-basic block basis, as follows. 
For each basic block $BB_i$, we start by:
i) randomly choosing a new IV $IV_i$;
ii) writing $IV_i$ at $BB_i$'s location in memory; and
iii) initialising the cipher state with $\mathcal{I}_{k}(IV_i)$. These actions are performed through a single \code{initBB} instruction.
%

Then, for each instruction making up $BB_i$, we:
i) write the encoding of the instruction into a general purpose register;
ii) encrypt the content of this register by feeding it into $\mathcal{T}_{enc, \: k}$, using the \code{enc\_word} instruction; and
iii) write this content back to memory.


\subsubsection{Forward Jumps}

Special attention is needed when generating code for jump instructions, as consideration must be given to the position of the target address compared to the address of the jump. 
Consider a jump instruction \code{j $@t$}, targeting address $@t$, that we want to generate at address $@j$.
This generation is performed by the macro \code{JUMP($@t$)} in the \polen{} runtime library.

In the case of a backward jump, i.e. $@t<@j$, the target instruction has already been generated and $@t$ is known. 
The jump instruction can thus be produced immediately when \code{JUMP($@t$)} is called.

In the case of a forward jump, i.e. $@t>@j$, the target instruction has not yet been generated. 
Because noise instructions can be later added (code between $@j$ and $@t$), the value of $@t$ cannot be known.
When called, \code{JUMP($@t$)} encrypts the word with value \code{0x0} with the current state of the cipher, writing this value $\mathcal{T}_{enc, \: k}(state_{@j}, \mathtt{0x0})$ to $@j$.
When execution of the SGPC is complete, $@t$ is known.
The encoding of the jump instruction, including the \textit{correct} $@t$ is computed and patched at location $@j$.
To do this, we use the property of the cipher shown in Equation~\ref{eq:homomorphic_like}.
More precisely, the value $x = \mathcal{T}_{enc, \: k}(state_{@j}, \mathtt{0x0})$ is first obtained from memory.
Then, the value is patched by computing $x_{patch} = x \odot @t$.
From Equation~\ref{eq:homomorphic_like}, we can verify that:
\begin{align*}
    x_{patch} &= \mathcal{T}_{enc, \: k}(state_{@j}, \mathtt{0x0}) \odot @t\\
        &= \mathcal{T}_{enc, \: k}(state_{@j}, \mathtt{0x0} \oplus @t) \\
        &= \mathcal{T}_{enc, \: k}(state_{@j}, @t)
\end{align*}
Finally, the value $x_{patch}$ is written back to memory at location $@j$.
%
This technique is used for all types of jumps (relative, absolute, conditional), and implemented in the dedicated \code{CDGEND} function in the \polen{} runtime library.

\subsubsection{The \polen{} Runtime API With Code Encryption}

The SGPC function that produces encrypted instances is adapted as shown in Figure~\ref{fig:enc SGPC}.
It is important to note that this code is generated by the static compiler, with no intervention from the programmer. 
A key difference with the version presented in Section~\ref{sec:code polymorphism} is that each basic block now includes an initialisation phase, implemented using the \code{initBB} instruction presented above. 
Furthermore, the generation of encrypted instructions is encapsulated into encoding macros, e.g., \code{ADD\_i\_1\_32\_iRRR(a0,a1,a0)} and now incorporates the three steps described in Section~\ref{sec:encrypted instruction generation}. 
Figure~\ref{fig:enc poly instance} shows an encrypted instance produced by \code{SGPC\_f\_secure}.
Each instruction is encrypted in memory and a comment is added in the Figure to give the corresponding decrypted instruction, for readability.

\begin{figure}
  \centering
  \subfloat[SGPC used in the example given in Figure~\ref{fig:example code cogito}.
  \label{fig:enc SGPC}] {
    \begin{minipage}[t]{.45\linewidth}
\begin{lstlisting}[frame=single,basicstyle=\ttfamily\scriptsize,language=C,morekeywords={INIT_BB,CDGEND}]
void SGPC_f_secure(){
  reg_t r[] = {0, 1, 2, 3, .... 15};
  shuffle_regs(r);
  push_callee_saved_registers();

  start_of_bb = (int*)cdg_asm_pc;
  __asm__ __volatile__ ("mv t4, %0" 
			: 
			: "r"(start_of_bb) 
			: "t4");
  INIT_BB();
  
  cdg_gennoise();
  variant_XOR_i_1_32_iRRR( a2, a1, a0 ); 
  cdg_gennoise();
  ADD_i_1_32_iRRR( a0, a1, a0 ); 
  cdg_gennoise();
  REM_i_1_32_iRRR( a0, a0, a2 ); 
  cdg_gennoise();
  pop_callee_saved_registers();
  cdg_gennoise();
  JALR_i_1_32_iIRR( zero, ra, 0 );
  ...
  CDGEND();
}

\end{lstlisting}
    \end{minipage}
  }
  \qquad
  \subfloat[Code (memory dump) of a polymorphic encrypted instance, for the example given in Figure~\ref{fig:example code cogito}. All instructions are encrypted in memory. Comments are added manually for reference.\label{fig:enc poly instance}]
  {
    \begin{minipage}[t]{.45\linewidth}
\begin{lstlisting}[frame=single,language={[riscv]Assembler}]
<code_f>:
  0x4bb535d6   ; IV1
  0x17a3e975   ; IV2
  0x2885fa3d   ; IV3

  0xff64fdac   ; encrypted push a5,a7,a8,a9
  0xf12f3809   ; encrypted sem variant for ...
  0xb5b21433   ; ... xor a2, a1, a0
  0x799f4c1a   ; ... cont'd
  0x6cdaa8e6   ; encrypted add a0,a1,a0
  0xdacd347d   ; encrypted noise sub a9, #127  
  0x4901978a   ; encrypted rem a0,a0,a2
  0xb3b5f0f9   ; encrypted noise add a7,a9,#5
  0xec171ee3   ; encrypted noise xor a8,#3
  0xd7c0b5ca   ; encrypted pop a5,a7,a8,a9
  0x5c84f0b1   ; encrypted ret
\end{lstlisting}      
    \end{minipage}
  }
  \caption{Dynamic code generation.}
  \label{fig:polen output}
\end{figure}

\subsection{\polen{} Configurability}
\label{sec:configurability}


One key aspect of \polen{} is its high degree of configurability, which is due to the following elements. 
First, code encryption and code polymorphism can be activated independently.
This creates four high-level configurations, labelled \plainC, \encC, \polyC{} and \polenC. 
In setting \plainC{}, no countermeasure is applied, while in setting \encC{} (resp.\ \polyC) only code encryption (resp.\  code polymorphism) is activated.
In setting \polenC{}, both code encryption and code polymorphism are activated.
Moreover, the programmer can choose to encrypt: i)~only the polymorphic instance; ii)~only the SGPC and wrapper; or iii)~both.
With respect to code polymorphism, the developer can activate each form of variability (noise, semantic variants, instruction shuffling, register shuffling) independently. 
In particular, for noise, he or she can also configure the amount of noise, and its probability model separately.
Finally, the instance regeneration frequency is configurable.
This gives the programmer great flexibility in trade-offs between security and performance and the ability to adapt \polen{} to the application context and corresponding threat model.

\section{Proof of Concept and Experimental Evaluation}
\label{sec:evaluation}

\dmr{We have implemented a proof of concept implementation of \polen{} based on the RISCV architecture and the LLVM compiler.
Section~\ref{sec:setup} describes this setup. It also discusses the implementation of the encryption/decryption modules as well as the corresponding hardware configurations we have considered. Finally, it describes the construction of simulated side-channel traces needed for our security evaluation.
Sections~\ref{sec:sec analysis} and \ref{subs:performance} give a security and a performance evaluation of \polen{}, respectively.}

\subsection{\dmr{Implementation and Experimental Setup}}
\label{sec:setup}
\paragraph{Simulation}
All our hardware modifications have been implemented as extensions to the \spike{}  Instruction Set Simulator,
a functional simulator that supports both 32- and 64-bit base ISAs, with multiple extensions~\cite{spike2017}.
\polen{} is based on the single core RV32IM ISA~\cite{WatermanRISCVInstructionSet2019} (I for Integer and M for Multiplication).

\paragraph{Toolchain}

The \polen{} toolchain comprises a set of passes that are added to the LLVM compiler~\cite{Lattner2004}, at both  backend and middle-end levels.
\polen{} is based on LLVM version 7.0.0 and \spike{}  1.0.1-dev implementing the RISC-V 
\dmr{Unprivileged ISA v2.0.}
All library, application and secured code is compiled in {\tt -O2}, but all optimisation levels are supported.
%
%

\paragraph{Choice of cipher}

So far, \polen{} has been presented as an abstract representation of the underlying encryption primitive introduced in Section~\ref{sec:encryption_formalism}.
Decryption is performed on a critical path of the processor, namely, in the \textsc{Fetch} stage.
An ideal cryptographic primitive would have a small footprint to limit the impact on the existing logic, and at the same time have a low latency that would not impact the processor's critical path.

\dmr{Given these constraints, we choose the Trivium~\cite{estream} stream cipher for the implementation of encryption and decryption modules. 
It is both lightweight and has low latency~\cite{rogawski2007hardware}. 
The number of bits decrypted per-cycle can be configured, which allows a trade-off between area and performance.
It should be noted that Trivium uses 80-bit keys, which is low considering current key size standards and recommendations.
However, to the best of our knowledge, it has not been subject to any significant attacks.
That being said, Trivium can be replaced by any other stream cipher in \polen{}, as shown in the next paragraph. Section~\ref{sec:block ciphers} also discusses the use of block ciphers.}

Stream ciphers (such as Trivium) work by generating a pseudorandom digit stream that is XOR-ed with the input message to produce the ciphertext.
Usually, they are built around two functions: 
\code{init} generates an initial, random-looking state, while \code{update} produces a new state and generates random output. Thus, with respect to the framework presented in Section~\ref{sec:encryption_formalism}, a stream cipher is expressed as:
\begin{itemize}
    \item $\mathcal{I}_k(IV) = \code{init}(k, IV)$
    \item $\mathcal{T}_{enc, \: k}(state, m) = (state',\: output \oplus m)$, where $(state',\: output) = \code{update}(state)$
    \item $\mathcal{T}_{dec, \: k}(state, c) = \mathcal{T}_{enc, \: k}(state, c)$
\end{itemize}
With a stream cipher, $\mathcal{I}_k(IV)$ is usually a slow operation, while $\mathcal{T}_{enc, \: k}$ and $\mathcal{T}_{dec, \: k}$ are fast.
\dmr{As $\mathcal{I}_k(IV)$ is only called when a control-flow instruction is taken, most of the time, we expect to be running the fast operations $\mathcal{T}_{enc, \: k}$ and $\mathcal{T}_{dec, \: k}$.}

\paragraph{Evaluated configurations}
Our evaluations are performed on the configurations listed in Table~\ref{tab:configurations}.
These six configurations are based on the activation (or not) of code
polymorphism and code encryption, respectively.

Concerning code polymorphism, the insertion of noise instruction is defined by a simple probabilistic model as follows.
Noise instructions are inserted in between two original instructions with a probability of $2^{-p}$, and
the number of noise instructions inserted is $2^N$, where $N$ is a uniform integer variable in $[1; N_{max}]$.
All of our evaluations are run with noise \code{nop} instructions only, $p=3$, and $N_{max}=5$.

With respect to code encryption, we evaluate two configurations of the cipher, denoted *\_9 and *\_35.
These correspond to hardware configurations where the initialisation of the cipher takes 9 (resp. 35) CPU cycles.
When there is no need to distinguish between *\_9 and *\_35, we note this as \encC{} or \polenC{}.
Configurations *\_35 are obtained with a Trivium instance that generates 32 bits of pseudorandom stream per clock cycle, which is the smallest configuration in terms of hardware area that can keep up with the CPU's instruction and data throughput.
%
%
The *\_9 configurations are obtained with a Trivium instance that generates 128 bits of pseudorandom stream per clock cycle.
Although these configurations require more hardware resources, the re-initialisation of the cipher (the function $\mathcal{I}_k$) is much faster. While adding more bits per clock cycle is possible, it is not worth the hardware overhead~\cite{Hiscock2019}. Thus, *\_35 (resp. *\_9) correspond to an \emph{area optimised} (resp. \emph{performance optimised}) version of the countermeasure. 

In our evaluation setting, when encryption is combined with code polymorphism, for each secured function, code encryption is applied to the wrapper, the SGPC and the polymorphic instances.

\newcommand{\listingsymbol}{\on\thinspace}
\begin{table}
  \caption{\polen{} configurations used in the evaluation.}
  \label{tab:configurations}
  \centering
  \begin{tabular}{|c|c|c|c|c|}\hline
    \multirow{2}{*}{configuration} & code polymorphism & \multicolumn{3}{c|}{encryption}                                       \\\cline{2-5}
                                   & activated                            & activated & target   & re-init cost (in $nb_I$) \\\hline
    \plainC                          & \off & \off      & --       & --                     \\\hline
    \encC\_9                         & \off & \on       & \listingsymbol function & 9       \\\hline
    \encC\_35                        & \off & \on       & \listingsymbol function & 35      \\\hline
    \polyC                           & \on  & \off      & --       & --                     \\\hline
    \polenC\_9                       & \on  & \on       & 
                                                         \begin{tabular}{@{\listingsymbol}l@{}}
                                                           wrapper                                                                               \\
                                                           SGPC                                                                             \\
                                                           instance                                                                              \\
                                                         \end{tabular}
                & 9                                                                                                                              \\\hline
    \polenC\_35 & \on & \on & 
                                                          \begin{tabular}{@{\listingsymbol}l@{}}
                                                            wrapper                                                                              \\
                                                            SGPC                                                                            \\
                                                            instance                                                                             \\
                                                          \end{tabular}
                                   & 35                                                                                                          \\\hline
  \end{tabular}
\end{table}

\paragraph{Production of simulated side-channel traces}
\label{sec:Production of simulated side-channel traces}

We obtain side-channel traces with the \spike{} simulator.
These traces include information for each instruction executed.
The RV32IM ISA includes the following four types of instructions: R, I, S and U.
Each has a different format and, of particular interest for us, addresses different registers (see Figure~\ref{fig:encoding}, left side).
Samples recorded into traces contain five, 32-bit integer values \code{<PC,INSN,r0,r1,r2>} where: \code{PC} is the current value of the Program Counter (the address of the current instruction); \code{INSN} is the content of the instruction register (i.e. the encoding of the currently-executed instruction, as decrypted by the $\mathcal{T}_{dec}$ module); and \code{r0}, \code{r1} and \code{r2} are values of registers (Figure~\ref{fig:encoding}, right side), depending on the instruction type. 
These values are recorded after the instruction has fully propagated to the CPU and memory.
One sample is generated for each instruction execution.
An exception is the \code{initBB} instruction, for which we generate three samples.
These correspond to the three CPU cycles associated with the memory reads required to obtain the IV value.
We do not differentiate configurations with different initialisation costs (Table~\ref{tab:configurations}) as Trivium's Initialisation Vector is a public value; hence, initialisation is completely independent of the target's secret values, and is not considered in our side-channel analysis.
It should be noted that this does not impact our security analysis and reduces the overall size of traces.

\begin{figure*}
  \centering\small
  \begin{tabular}{cccccccccccccccccc}
    31 & 27 & 26 & 25 & 24 & 20 & 19 & 15 & 14 & 12 & 11 & 7 & 6 & 0 & & \code{r0} & \code{r1}  & \code{r2} \\\cline{1-14}\cline{16-18}
    \multicolumn{4}{|c|}{funct7} & \multicolumn{2}{c|}{rs2} & \multicolumn{2}{c|}{rs1} & \multicolumn{2}{c|}{funct3} & \multicolumn{2}{c|}{rd} & \multicolumn{2}{c|}{opcode} & R-type & \multicolumn{1}{|c|}{rs1} & \multicolumn{1}{|c|}{rs2} & \multicolumn{1}{|c|}{rd}\\\cline{1-14}\cline{16-18}
    
    \multicolumn{6}{|c|}{imm[11:0]} & \multicolumn{2}{c|}{rs1} & \multicolumn{2}{c|}{funct3} & \multicolumn{2}{c|}{rd} & \multicolumn{2}{c|}{opcode} & I-type & \multicolumn{1}{|c|}{rs1} & \multicolumn{1}{|c|}{0} & \multicolumn{1}{|c|}{rd} \\\cline{1-14}\cline{16-18}
    
    \multicolumn{4}{|c|}{imm[11:5]} & \multicolumn{2}{c|}{rs2} & \multicolumn{2}{c|}{rs1} & \multicolumn{2}{c|}{funct3} & \multicolumn{2}{c|}{imm[4:0]} & \multicolumn{2}{c|}{opcode} &  S-type &   \multicolumn{1}{|c|}{rs1} & \multicolumn{1}{|c|}{rs2} & \multicolumn{1}{|c|}{rd} \\\cline{1-14}\cline{16-18}

    \multicolumn{10}{|c|}{imm[31:12]} & \multicolumn{2}{c|}{rd} & \multicolumn{2}{c|}{opcode} & U-type & \multicolumn{1}{|c|}{0} & \multicolumn{1}{|c|}{0} & \multicolumn{1}{|c|}{rd} \\\cline{1-14}\cline{16-18}
\end{tabular}

\caption{RV32-IM instruction encoding formats (left). Sample content, as a function of the instruction type. Samples are of the form \code{<PC,INSN,r0,r1,r2>} (right).}
\label{fig:encoding}
\end{figure*}

\subsection{Security Analysis}
\label{sec:sec analysis}

This section presents a security evaluation of an AES implementation.
This implementation is the same as the AES 8-bit implementation evaluated in Section~\ref{subs:performance} (\aesb).
Our analysis focuses on an implementation of AES as this cryptographic primitive is widely used in many embedded systems, ranging from IoT devices to mobile and desktop computers, and its original implementation does not include any side-channel attack countermeasures.
The security against side-channel attacks of the AES cipher has been studied for several years, and it is often used as a reference for security evaluation.

We base our analysis on simulation traces obtained by instrumenting the \spike{} simulator (\autoref{sec:Production of simulated side-channel traces}). Since \spike{}  is a functional simulator and does not include details of the processor's microarchitecture, our traces are free from the side-effects, e.g., resulting from pipelined execution of instructions.
Our simulated traces include, for each CPU cycle, the value of the PC register and the binary encoding of the executed machine instruction.
It is common practice to assume that, on a processor without side-channel protection, addresses and values on instruction buses may leak.
Furthermore, we include the value of the destination and source registers, as recent research findings show that such microarchitecture elements may be vulnerable~\cite{BarenghiSidechannelsecuritysuperscalar2018}.
\dmr{Finally, traces are noise-free. In terms} of security, this provides a pessimistic bound on the results of the evaluation.
This issue is discussed in more detail later in this section.

The side-channel simulation generates traces of 32-bit unsigned integer values that represent internal data manipulated by the processor.

For the analysis, we use the Hamming Weight of 32-bit sample \dmr{values. This choice follows} standard practice which assumes that, in software implementations, leakage follows a Gaussian distribution centred on the Hamming Weight of the target secret value~\cite{MangardPoweranalysisattacks2007}.


\subsubsection{Leakage Analysis}
\label{sec:nicv}

\newcommand{\XXX}{\textbf{X}}
\newcommand{\sensVar}{z}
\newcommand{\sensRandVar}{Z}
\newcommand{\esper}[2][]{\underset{#1}{\mathbb{E}}\left[#2\right]}
\newcommand{\var}[1]{\mathbb{V}{\left( #1 \right) }}
\newcommand{\varOn}[2][]{\underset{#1}{\mathbb{V}}\left(#2\right)}
\newcommand{\NICV}{\mathrm{NICV}}

The signal-to-noise ratio (SNR) is commonly used as a tool to detect the presence of trivial information leakage in side-channel traces~\cite{MangardPoweranalysisattacks2007}.
It is also used to select regions of interest in order to reduce the computational complexity of attacks such as Correlation Power Analysis (CPA).
However, in the presence of noise-free leakage traces, the SNR computation will produce infinite values, which make it unsuitable for the purposes of our evaluation.
Instead, we use the Normalized Inter-Class Variance (NICV) metric.
\dmr{The NICV was initially introduced by \citeauthor{BhasinNICVNormalizedInterClass2013} as a metric agnostic to the leakage model, 
providing a conservative bound for any other leakage analysis based on a specific model~\cite{BhasinNICVNormalizedInterClass2013}.}
In other words, the use of the NICV provides a conservative security metric of the leakage analysis that a real-life attacker could carry out.

For each time sample \(t\), the NICV is estimated with the following formula:

\begin{eqnarray}
	\NICV[t] & \triangleq & \frac{\varOn[Z]{\esper{\XXX[t]|
	\sensRandVar = \sensVar}}}{\var{\XXX[t]}} \enspace ,
\end{eqnarray}
where \(\XXX[t]\) is a random, continuous variable denoting the side-channel observation measured for sample \(t\), \(\sensRandVar\) is a random discrete variable denoting the sensitive target, \(\mathbb{E}\) denotes the expectation and \(\mathbb{V}\) denotes the variance.
When the sample \(t\) contains no leakage information, \(\XXX[t]\) does not depend on \(\sensRandVar\) and the numerator is zero.
It should be noted that the amount of noise in the measured traces directly impacts global variance $\var{\XXX[t]}$: the higher the measurement noise, the lower the NICV value.
In the absence of measurement noise, as is the case in our experimental setup, global variance $\var{\XXX[t]}$ is only impacted by variability due to computations, i.e., variability due to the data that is processed and variability due to the possible presence of countermeasures.
It should also be noted that such a leakage analysis is sometimes called a \emph{vertical} analysis, because it is computed on a per-sample basis.
As a consequence, if information leakage is spread across different samples $t$, the measured NICV value is degraded, because of the spread.

\begin{figure}
  \centering
  \begin{minipage}[b]{.49\linewidth}
    \subfloat[NICV, \plainC{} version.\label{fig:nicv unprotected}]{
	\includegraphics[width=\textwidth]{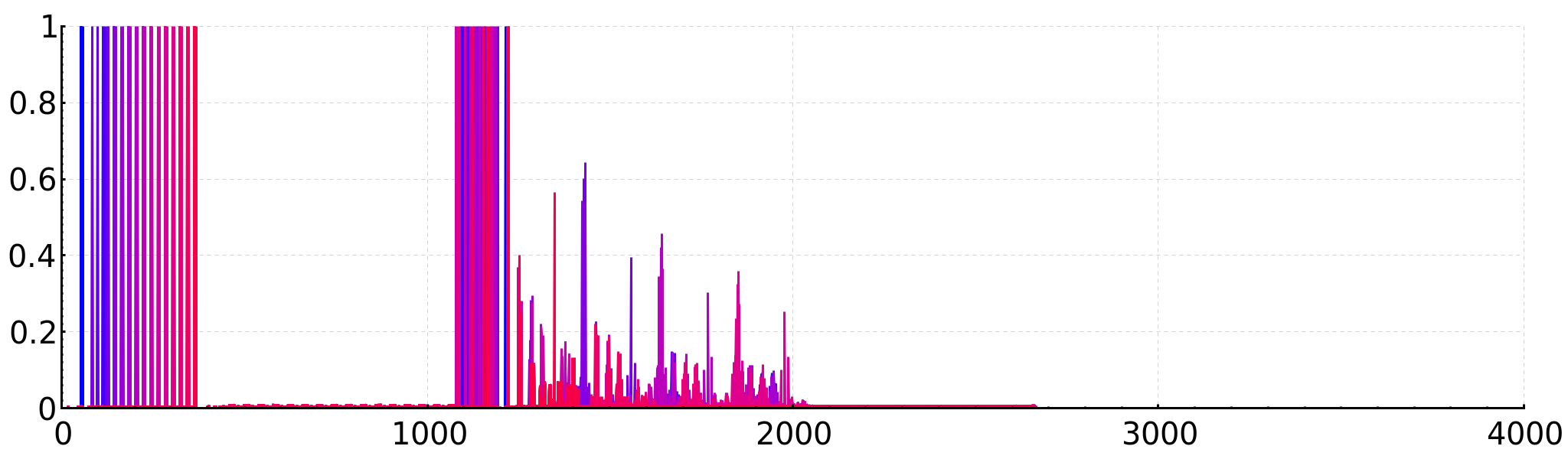}}
  \end{minipage}
  \hfill
  \begin{minipage}[b]{.49\linewidth}
    \subfloat[NICV, \encC{} version.\label{fig:nicv encrypted}]{
        \includegraphics[width=\textwidth]{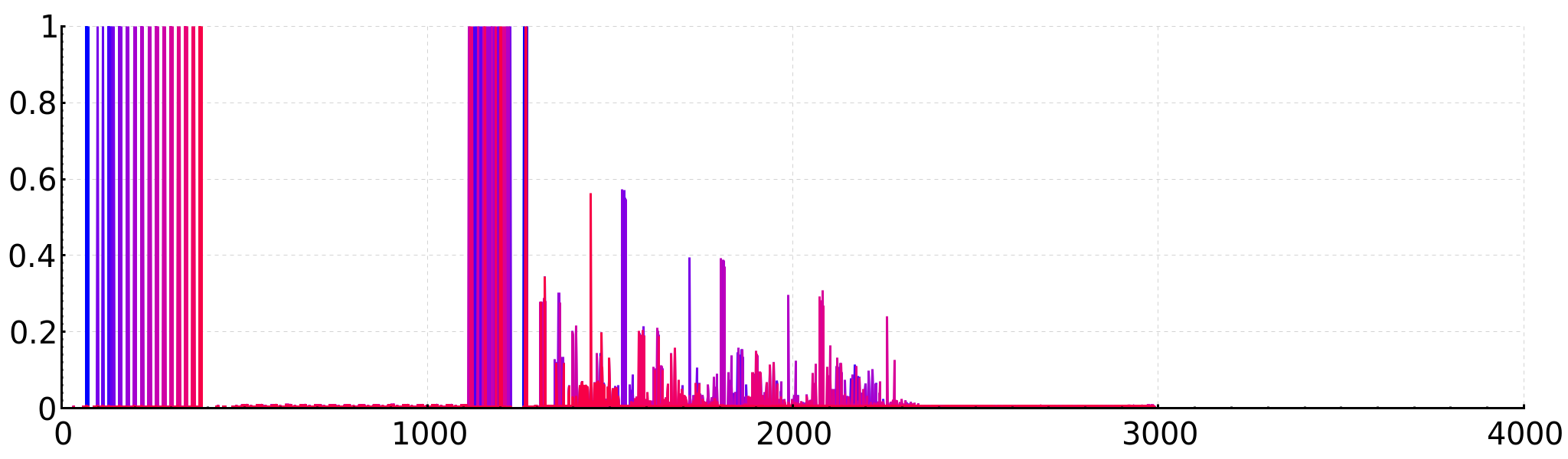}}
  \end{minipage}

    \begin{minipage}[b]{.49\linewidth}
    \subfloat[NICV, \polyC{} version.\label{fig:nicv polymorphic}]{
	\includegraphics[width=\textwidth]{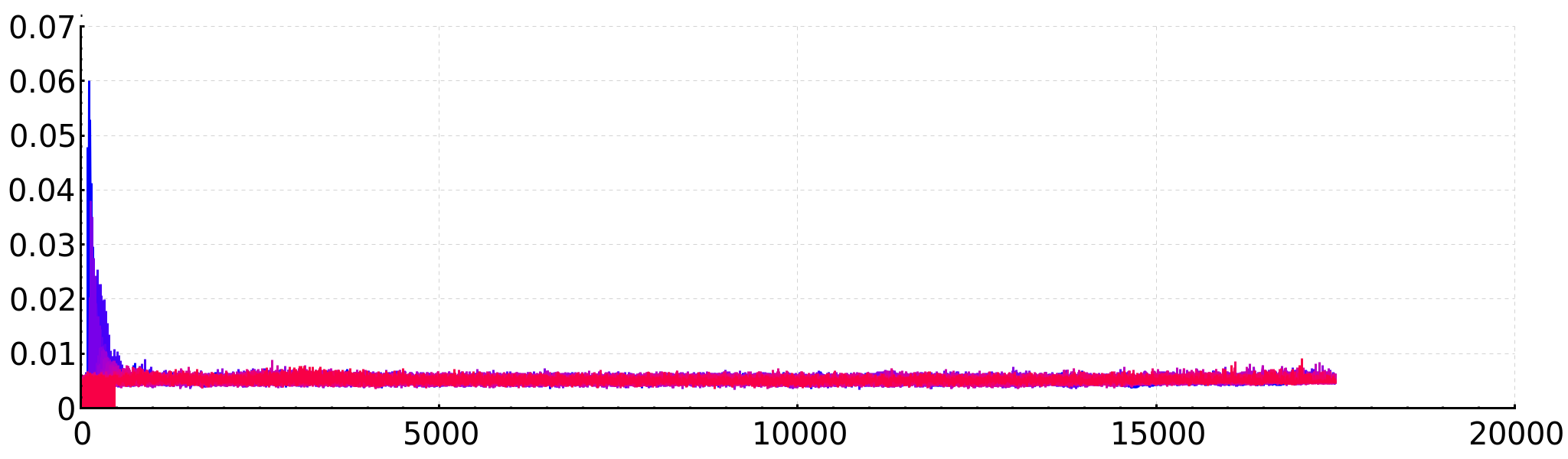}}
  \end{minipage}
  \hfill
  \begin{minipage}[b]{.49\linewidth}
    \subfloat[NICV, \polenC{} version.\label{fig:nicv polen}]{
        \includegraphics[width=\textwidth]{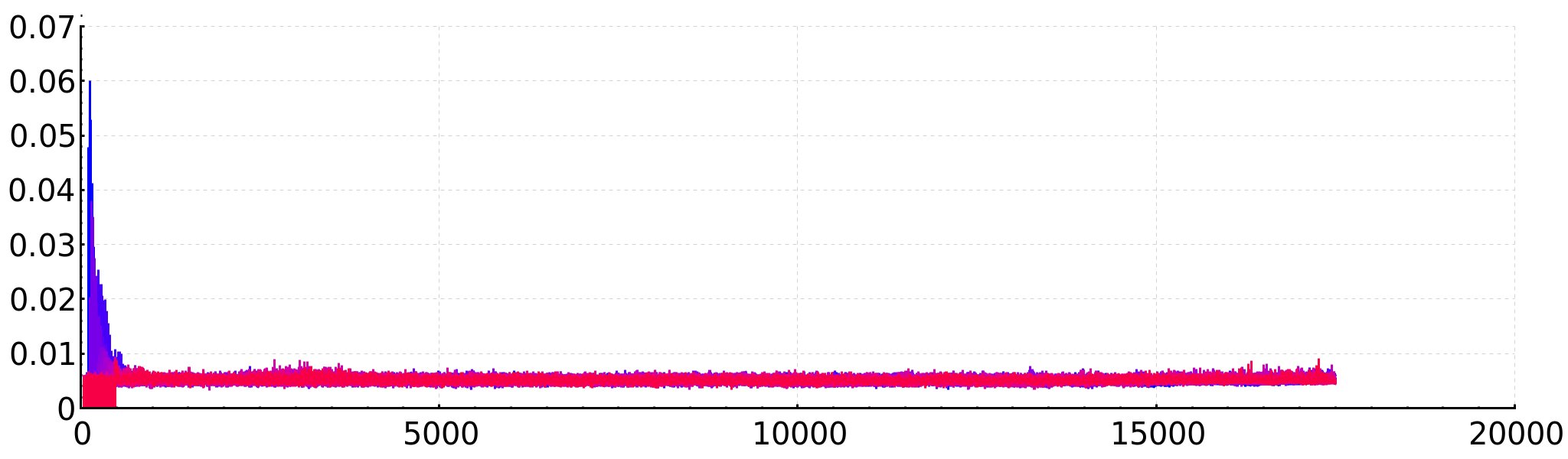}}
  \end{minipage}
  \caption{Results of the leakage analysis on 50,000 traces for all key bytes (ranging from byte 0 in blue, to byte 15 in red).}
  \label{fig:nicv}
\end{figure}

Figure~\ref{fig:nicv} illustrates the results of the leakage analysis for the four experimental configurations \plainC, \encC, \polyC{} and \polenC{}.
For \plainC{} and \encC, which are not protected against side-channel analysis, SNR values are 1.0 for all key bytes.
This result suggests strong information leakage, and we demonstrate below that this can be exploited by a chosen-plaintext attack (CPA).
Results for \encC{} illustrate that encrypted programs can still leak information in a side-channel analysis.

Results for the two polymorphic configurations, \polyC{} and \polenC, demonstrate a greatly reduced NICV,
to the point that it is impossible to identify significant points of interest for side-channel attacks.
Peaks can be identified at the very beginning of the NICV trace, for the least-significant key bytes (in blue).
However, these peaks are not correlated to secret values, and we verified that this potential leakage could not be exploited in a CPA (cf.\ Section~\ref{sec:cpa} and Figure~\ref{fig:cpa}).
In a configuration where the NICV does not exhibit strong information leakage, an attacker can still carry out a visual analysis of side-channel traces to identify trace features that could be exploited e.g., to re-align traces with more sophisticated processing methods.
However, this attack scenario is difficult to assess with objective metrics, and hence not considered in this study.
Another option for the attacker is to perform an exhaustive attack on the full set of samples available in the traces.
In this case, the quantity of data necessary for a successful attack depends on measurement conditions (notably, the amount of measurement noise and intrinsic noise due to parasitic switching activity in the target), which directly impact the computational complexity.
We evaluate the feasibility of this attack scenario below.

\subsubsection{CPA Analysis}
\label{sec:cpa}

We evaluated a first order CPA against the same four implementations. In this case, the attack targets the output of the first SubBytes function of the AES encryption.
%
Figure~\ref{fig:cpa} presents the results of the attack on the first key byte.
For each configuration, left plots illustrate the evolution of the correlation as a function of the number of traces, and right plots illustrate correlation curves for 500 and 100,000 traces for \plainC{} and \polyC{}, respectively.
An attack on the other key bytes leads to similar results; these results are not included for the sake of conciseness.
In \plainC{}, the absence of measurement noise in simulation traces means that the maximum correlation value for the correct key is around $0.5$, and this becomes distinguishable from the wrong key hypothesis with 100 traces.
Likewise, \encC{} is also vulnerable to a CPA.
This result was expected, since the CPA targets data manipulated by the program and, in our implementation, program encryption does not protect the data path of the processor.
In the case of measurements on a real circuit, the presence of noise degrades the effectiveness of a CPA (for example, noise due to the measurement setup and switching activity, independent of the targeted computations~\cite{MangardPoweranalysisattacks2007}).
Hence, our results provide a conservative bound on real security in a real attack scenario.
In the worst case, a secret key can be extracted from an unprotected circuit in a few dozen traces, compared to thousands or millions of traces in more complex situations.

Our results also illustrate the effectiveness of code polymorphism to protect the AES implementation against a CPA.
Figure~\ref{fig:cpa}(c--d) shows that the correlation value for the secret key cannot be distinguished from other key hypothesis up to approximately 80,000 traces for \polenC{}.
For \polyC{}, the secret key can be found with 100,000 \dmr{traces. However,} in Figure~\ref{fig:cpa}(c, left), the correlation of the correct key is undistinguishable from the other key hypotheses due to higher incorrect correlation peaks for samples in the trace window $[0; 500]$ (Figure~\ref{fig:cpa}(c, right)).
Our results are congruent with prior findings on similar software countermeasures~\cite{Agostacodemorphingmethodology2012,AgostaMEETApproachSecuring2015,Belleville2018}.

\newlength{\cpawidth}
\setlength{\cpawidth}{.48\textwidth}

\begin{figure}
  \centering
    \subfloat[CPA, \plainC{} version.\label{fig:cpa unprotected}]{
	\includegraphics[width=\cpawidth]{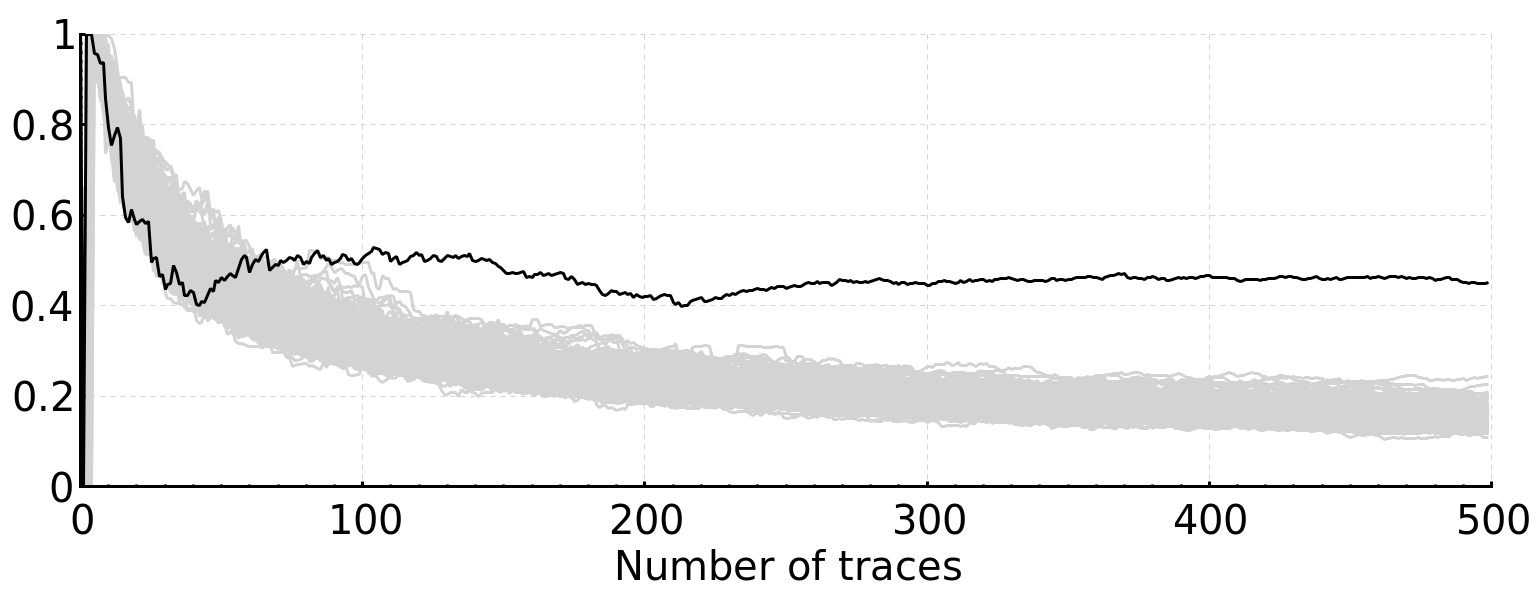}      \hspace*{\fill}
	\includegraphics[width=\cpawidth]{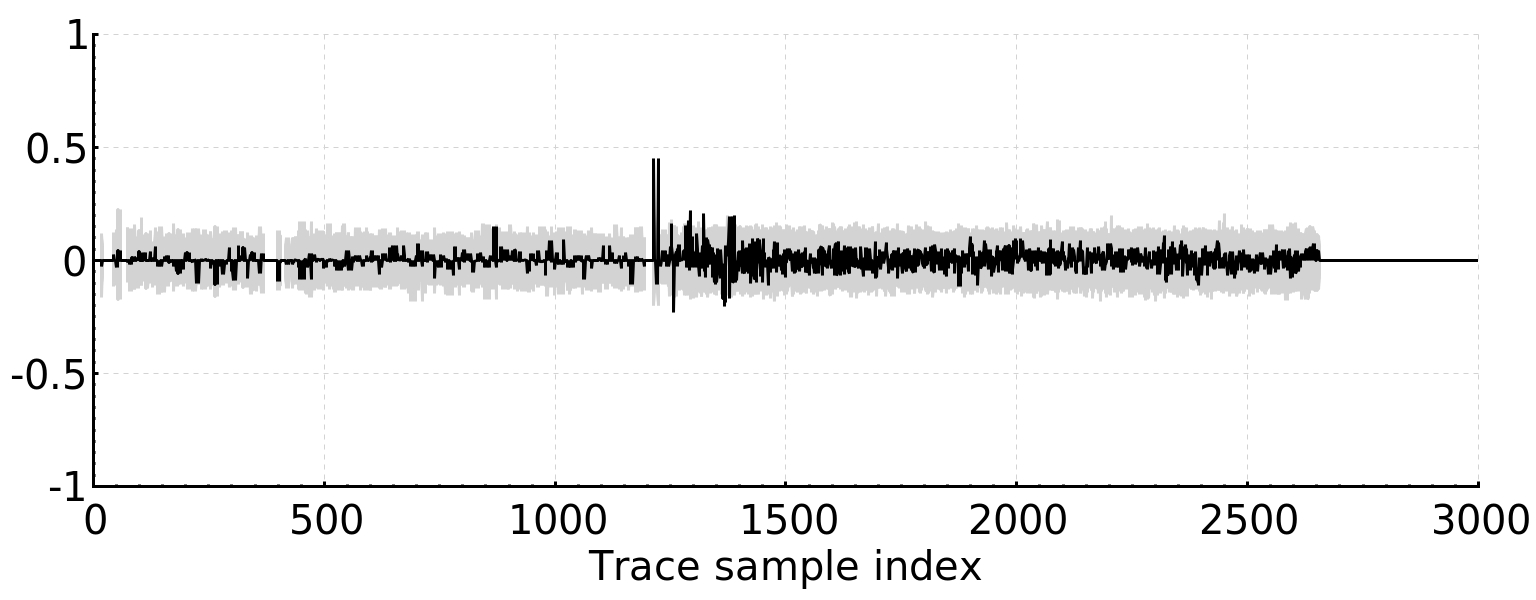}      }

    \subfloat[CPA, \encC{} version.\label{fig:cpa encrypted}]{
        \includegraphics[width=\cpawidth]{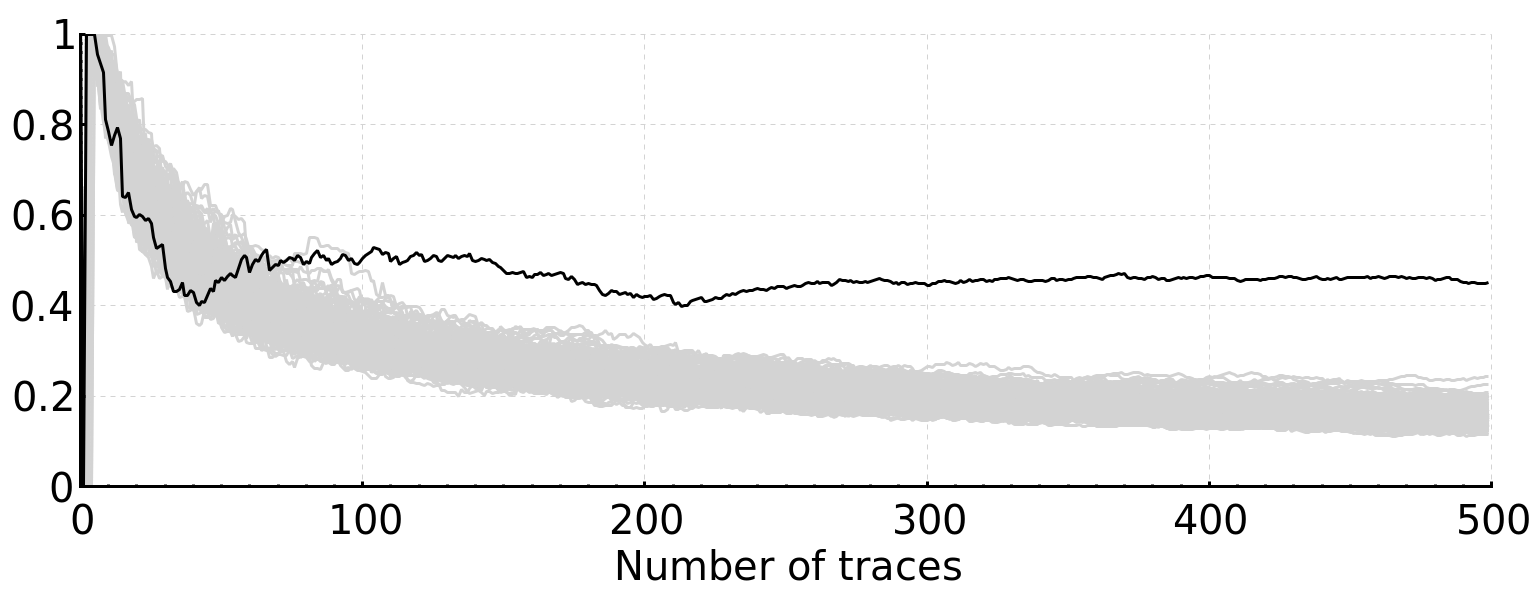}         \hspace*{\fill}
        \includegraphics[width=\cpawidth]{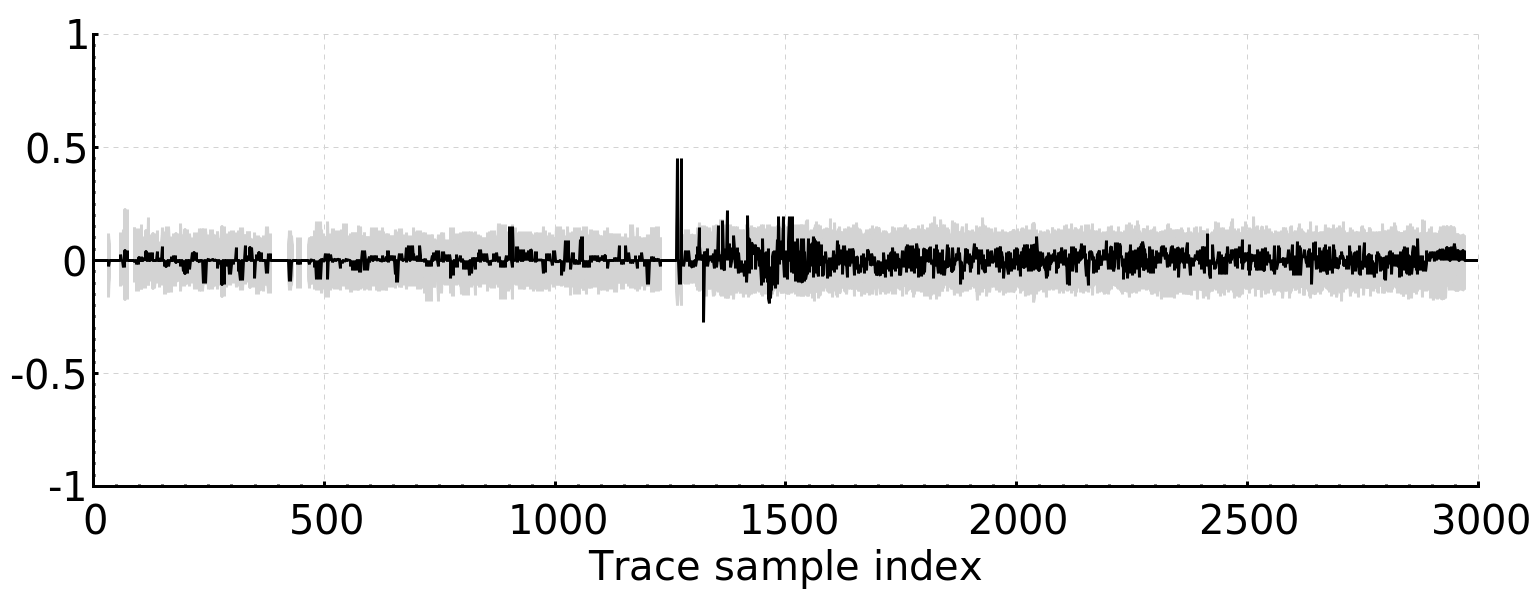}	   }

    \subfloat[CPA, \polyC{} version.\label{fig:cpa polymorphic}]{
        \includegraphics[width=\cpawidth]{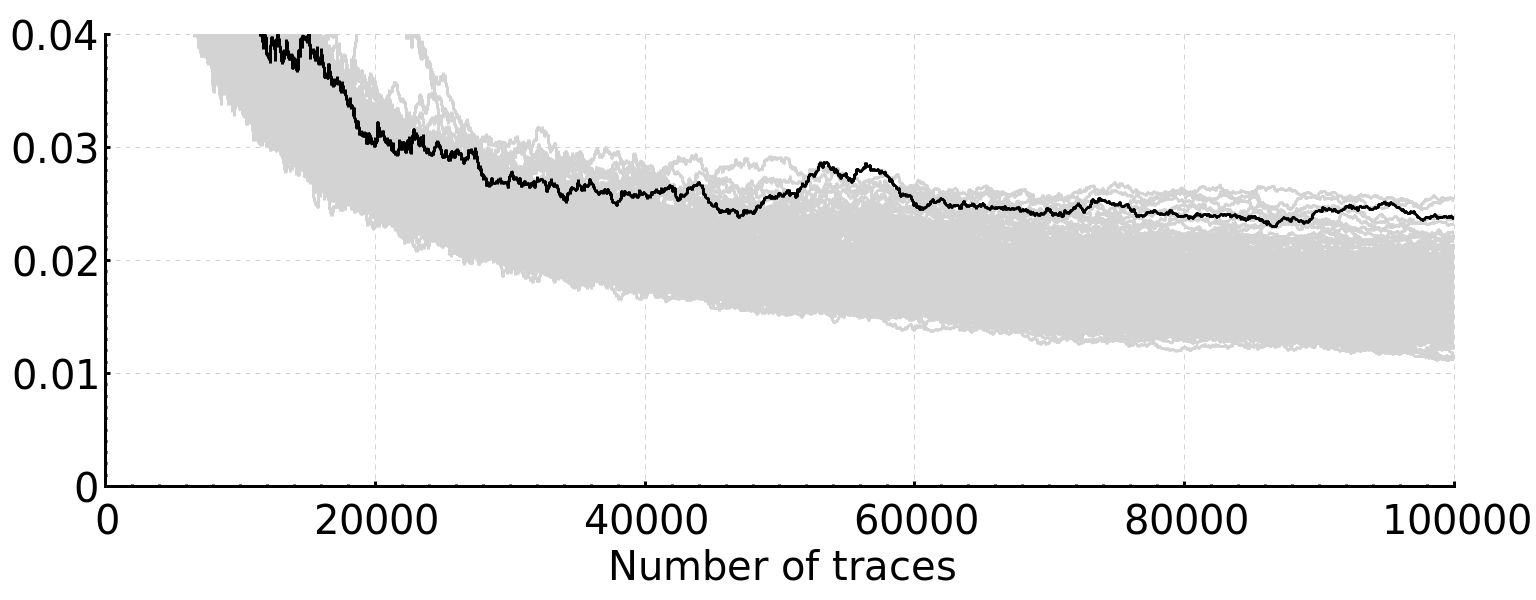}       \hspace*{\fill}
        \includegraphics[width=\cpawidth]{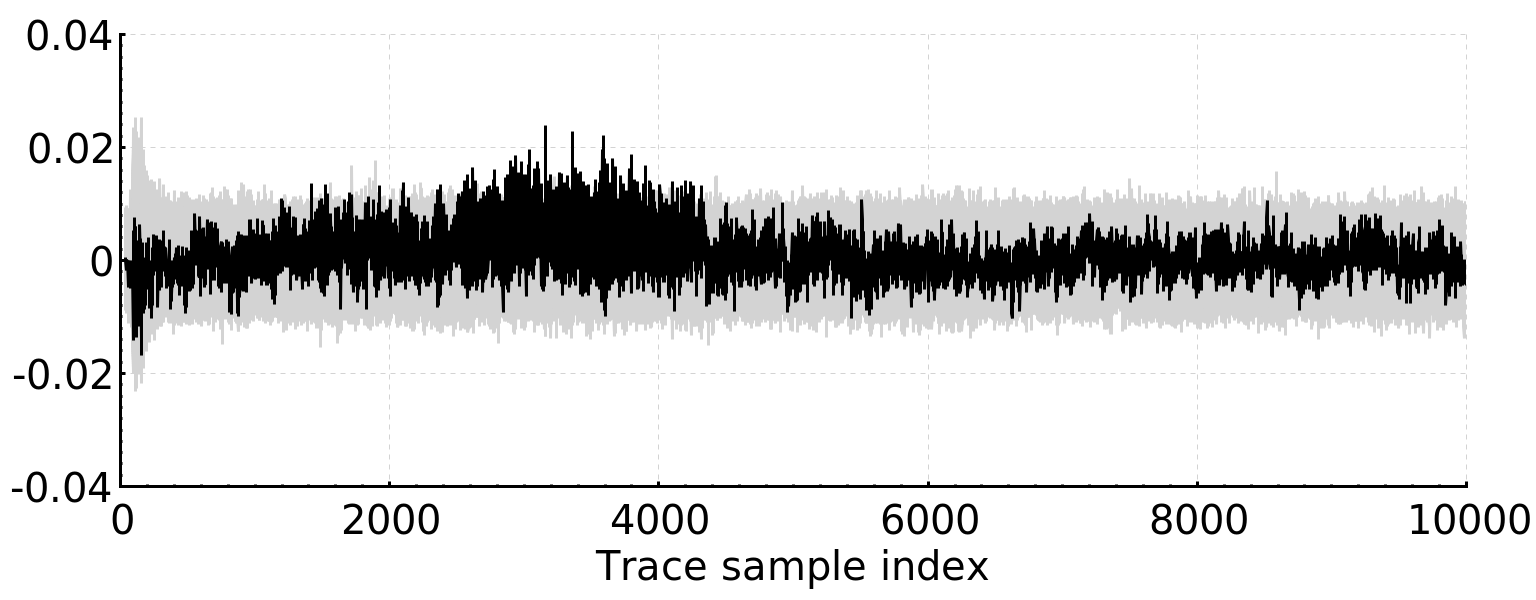}	}

    \subfloat[CPA, \polenC{} version.\label{fig:cpa polen}]{
        \includegraphics[width=\cpawidth]{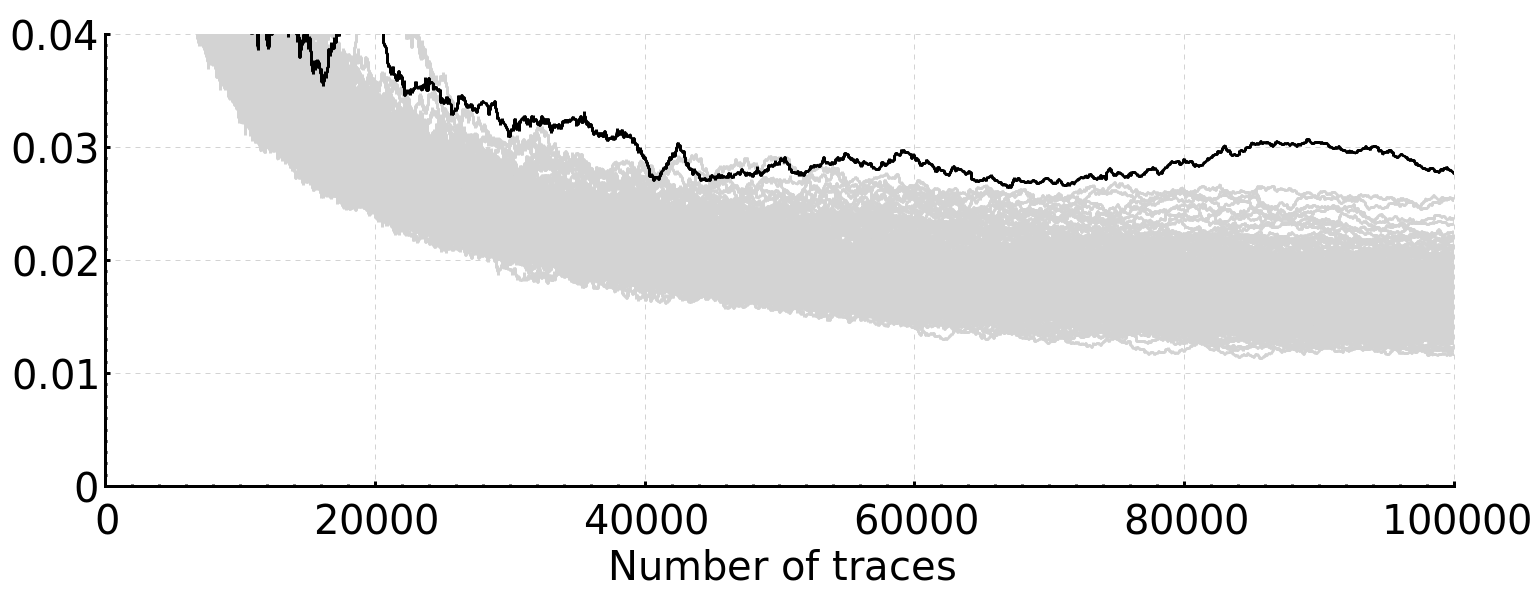}         \hspace*{\fill}
        \includegraphics[width=\cpawidth]{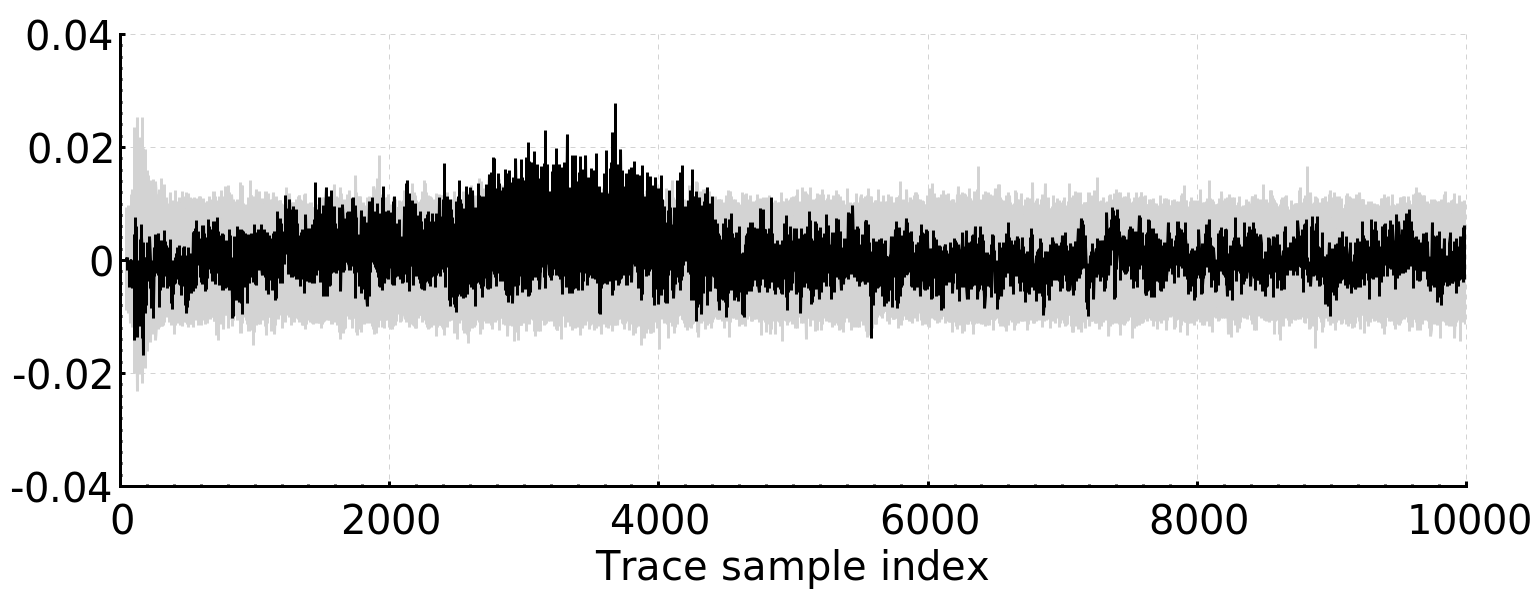}	}
      \caption{Results of the CPA analysis on 500 traces for \plainC{} and \encC, and on 100,000 traces for \polyC{} and \polenC.  The CPA targets the secret key byte 0 (analyses of the other key bytes give similar results).  The correlation curve is plotted in black for the correct key, and in grey for the wrong key hypothesis.}
  \label{fig:cpa}
\end{figure}

\subsection{Performance Analysis}
\label{subs:performance}

Performance is evaluated as a function of three main criteria: memory usage, execution time and hardware overhead.
The latter is discussed in Section~\ref{ssec: hw criteria}. 
We evaluate the first two using the nine programs listed in Table~\ref{tab:bench}. 
Five of these programs are taken from the mbed-TLS suite~\cite{mbedtls}, one is a software implementation of \trivium, taken from the estream suite~\cite{estream} and the last three are custom implementations of the \aesb, \misty~\cite{Matsui1997} and \simon~\cite{Beaulieu2013} protocols. 

Our results illustrate the impact of using the configurations listed in Table~\ref{tab:configurations} and presented in detail in Section~\ref{sec:setup}. The reader should keep in mind that all of our evaluations are based on the Spike simulator, which is \textit{not} cycle accurate. 
\begin{table}
\caption{Benchmarks used for the \polen{} evaluation.}
\label{tab:bench}
  \centering
  \begin{tabular}{|l|l|l|}\hline
  Nature        & Name of the benchmark           & Origin                  \\\hline
  block cipher  & \aesttable, \camellia, \des & mbed-TLS \cite{mbedtls} \\\hline
  block cipher              & \aesb, \misty, \simon       & custom                  \\\hline
  stream cipher & \trivium                    & estream \cite{estream}  \\\hline
  hash function & \md, \sha                   & mbed-TLS \cite{mbedtls} \\\hline
\end{tabular}
\end{table}

\subsubsection{Memory Usage}

We distinguish between statically and dynamically allocated memory. Static memory includes the \polen{} runtime library and the code generated statically (the wrapper and the SGPC) for each secured function. Dynamically-allocated memory corresponds to the space needed to hold polymorphic instances. We discuss both types of memory usage in the following, starting with static memory. 

Currently, the \polen{} library's size is 9.4~kB, which is suitable for low-memory on-chip embedded systems. This version was used in all evaluations and is not considered as a factor in analyses. 

\begin{figure}[t]
  \centering
\subfloat[Overhead related to the size of the object file generated for \encC, \polyC{} and \polenC{}.
\label{fig:overhead size objfile all vs plain}]{
\includegraphics[width=0.45\textwidth]{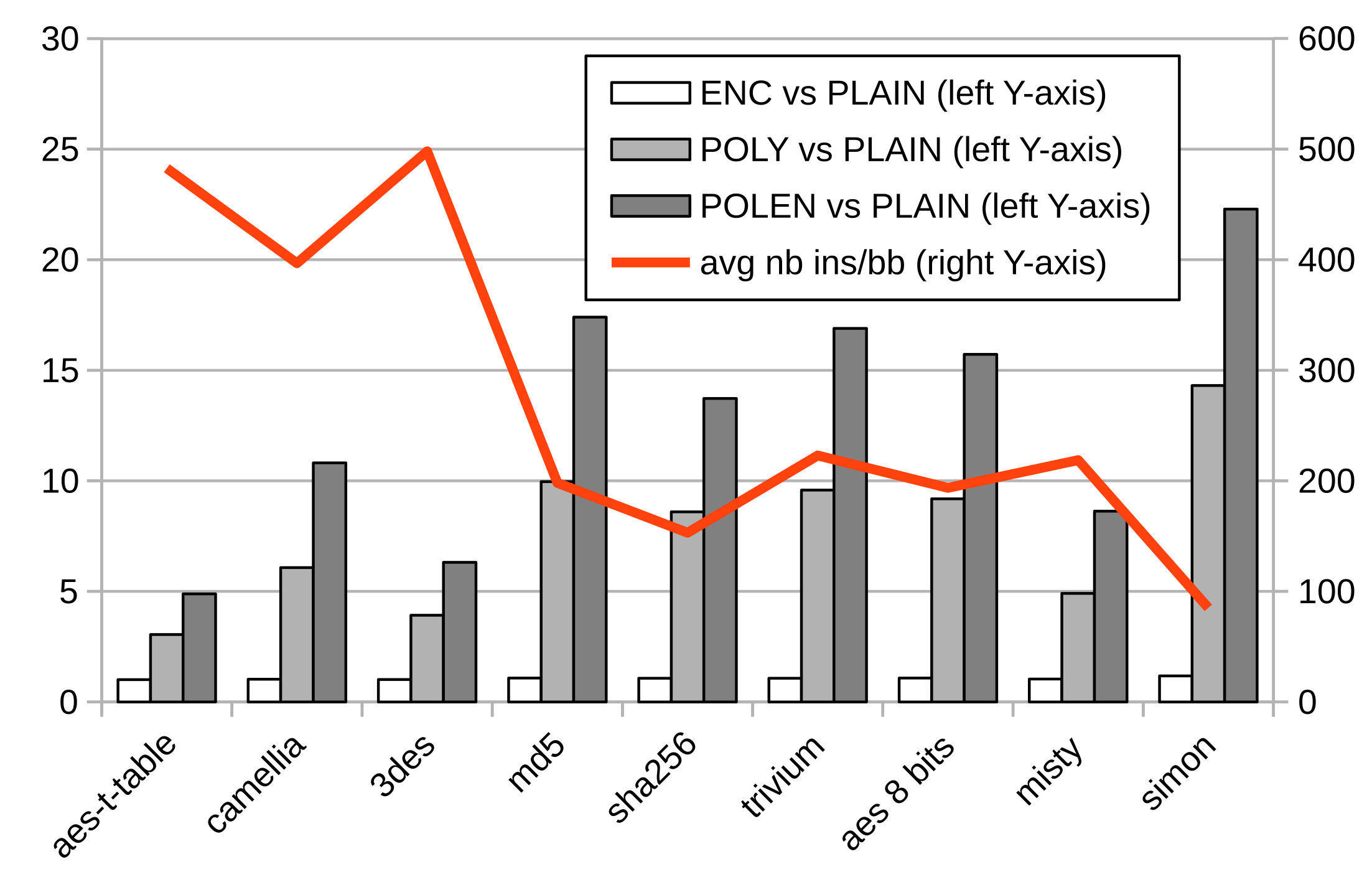}
}
\qquad
  \subfloat[Overhead related to dynamic memory used to generate polymorphic instances for \polyC{} and \polenC{}. 
  \label{fig:overhead size poly instance poly polen vs plain}]{
\includegraphics[width=0.45\textwidth]{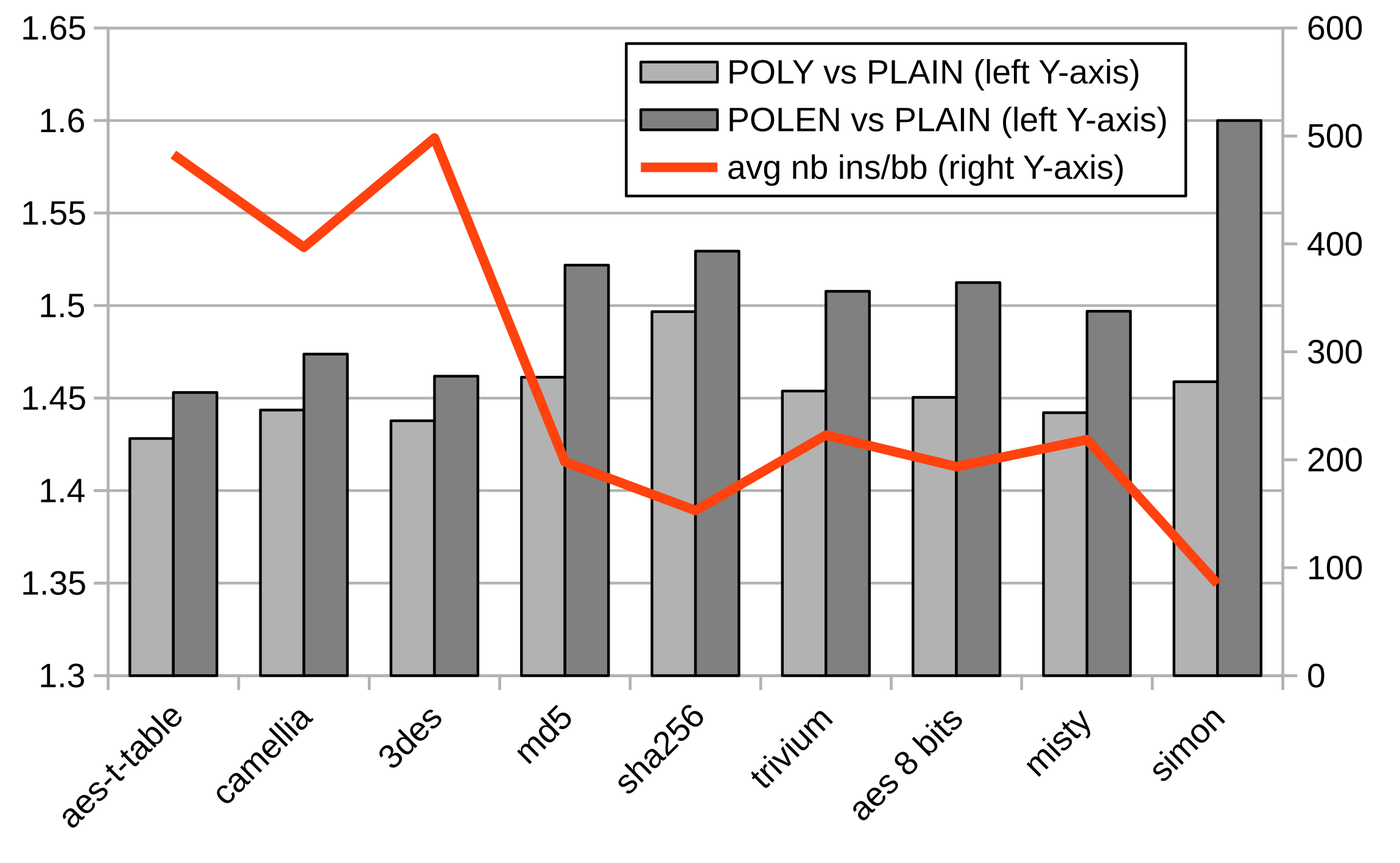}
    }
  \caption{Memory usage of \polen,
   expressed as a factor of the size of the object file for \plainC{} (left Y-axis). The red line shows the average number of instructions per basic block (right Y-axis).}
  \label{fig:mem overhead}
\end{figure}

For each program in our test suite, we measure the size of the object file containing the code generated for the secured function.
%
%
In Figure~\ref{fig:overhead size objfile all vs plain}, we report the overhead of this object file compared to the \plainC{} version, for \encC, \polyC{} and \polenC{}. For \encC, the overhead simply consists of the IVs that are added at the beginning of each basic block in the secured function. On average, this overhead is inversely proportional to the number of instructions in each basic block, shown as a red line. 
We added a constant-size IV to each basic block, regardless of its size. Consequently, the bigger the basic blocks, the better it compensates for the introduction of IVs.
In both \polyC{} and \polenC{} cases, the same overhead is observed, in this case, amplified by the size of the wrapper and SGPC code that is now included. 

For \polenC{}, there is an overhead related to the size of the
statically-built binary code of the secured function, of the order of
5 to 22. This overhead includes the application of code polymorphism itself, and encryption of both the wrapper and the SPGC. Although this is considerable, it should be put in perspective, as \polen{} is only applied to a part of a larger code base, depending on the ratio of protected and unprotected code.

The footprint of protected applications varies at \dmr{runtime. This overhead depends on} the quantity of dynamic memory needed to generate polymorphic instances. It is a function of the structure of the protected program, notably the size of basic blocks, their number and, thus, the number of IVs.
Figure~\ref{fig:overhead size poly instance poly polen vs plain} illustrates this for each example by giving the overhead related to the size of polymorphic instances compared to the size of the statically-generated code in the \plainC{} version.
Both \polyC{} and \polenC{} versions are plotted. For \polenC{} (resp. \polyC{}), this overhead varies between a factor of 1.45 and 1.6 (resp. 1.42 and 1.5). This factor is highly influenced by the addition of IVs, as can be observed by comparing the red line (showing the average number of instructions per basic block) to bar graphs.
%
%
This point becomes clearer when comparing the \simon and \des protocols. For \simon, there are approximately 100 instructions per basic block, and the overhead is far bigger than for \des, which has slightly less than 500 instructions. Both variability in the source used to generate polymorphic instances, and the amount of noise instructions inserted highly influence this overhead. This obviously needs to be taken into account when addressing the security/performance compromise, which is specific to each application.

\subsubsection{Execution Time}

We now evaluate the impact of the different configurations of \polen{} on execution time.
It should be recalled that, as \spike{}  is not cycle accurate, all evaluations of execution times are measured as an equivalent of number of instructions executed, denoted $nb_I$. 
\dmr{%
However, our evaluation still provides a realistic approximation of the performance overheads on simple cores (in-order, single issue).
The study of \polen{} on more advanced processors (superscalar, out-of-order) is not in the scope of this paper.  
}

\begin{figure}[t]
  \centering
  \includegraphics[scale=1.4]{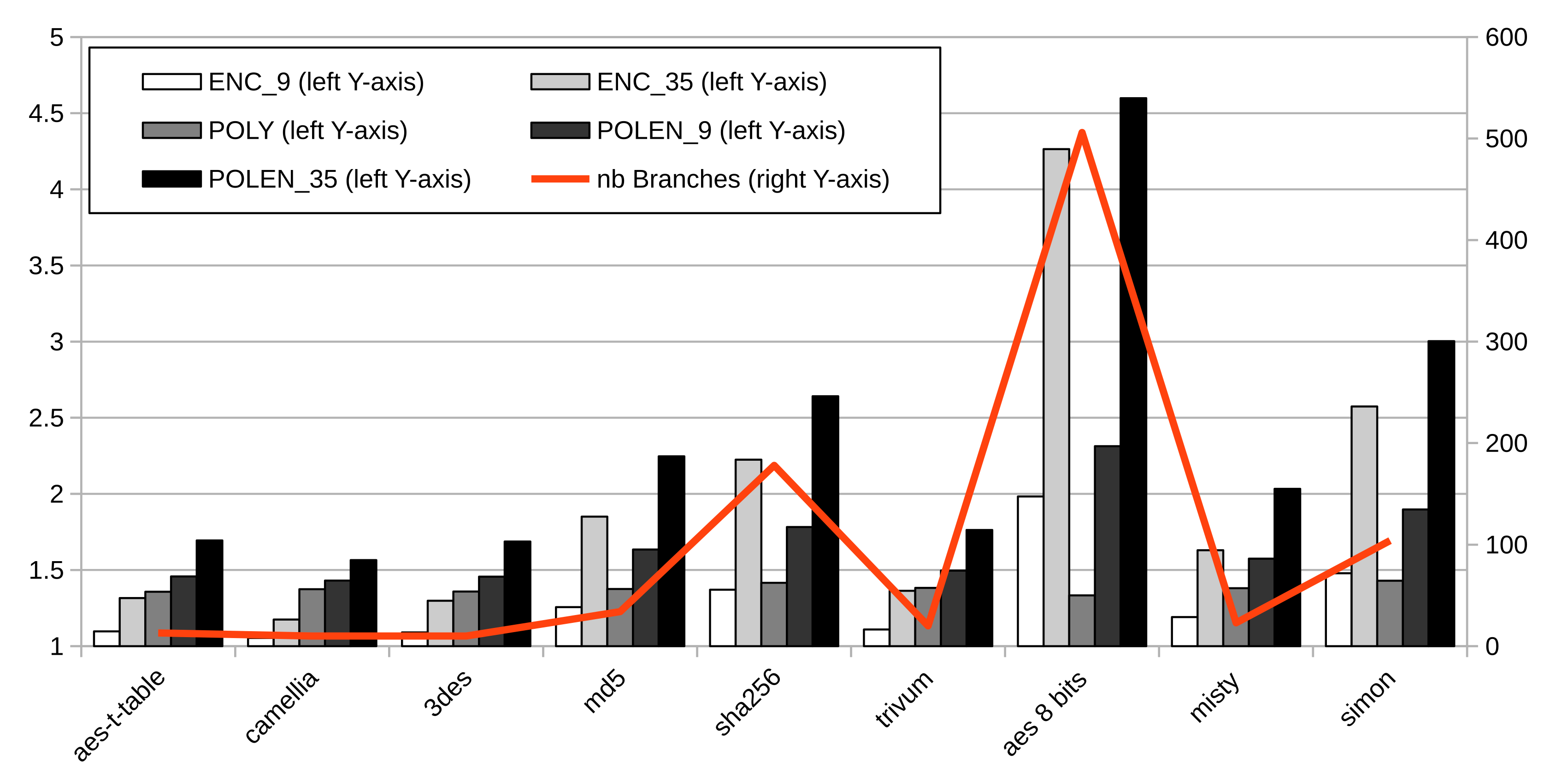}

  \caption{\dmr{Execution time overhead of the secured function for \encC{}, and polymorphic instances for \polyC{} and \polenC{}. Bars (indexed on the left Y-axis) represent overheads compared to the \plainC{}. The red line (indexed on the right Y-axis) gives the number of control-flow instructions taken during the execution of secured functions.}
  \label{fig:overhead execution time of polen instances}}
\end{figure}

First, we measure the overhead when executing a secure function compared to the \plainC{} version.
When code polymorphism is used, only the cost of executing a polymorphic instance is taken into account, and not the cost of generating the instance.
This is shown in Figure~\ref{fig:overhead execution time of polen instances}, which reports execution time overheads for configurations \encC\_9, \encC\_35, \polyC, \polenC\_9{} and \polenC\_35 as increasingly darker grey bars.
\dmr{The reported numbers are averaged over 100 executions of the secured function.
The graph also reports (the red line), the number of control-flow instructions taken during the execution of the \textit{unprotected} secured function.
This is an important point to note as taken control-flow instructions trigger the re-initialisation of the Trivium module and, thus, account for a major part of the execution time overhead. 
For polymorphic instances, the number of Trivium re-initialisations is identical for all configurations. 
For \encC{} and \polenC, the number of taken control-flow instructions directly impacts execution time, because each
of these instructions trigger a re-initialisation of the decryption module.
%
For \aesttable, \camellia, \des{} and \trivium, between 10 and 23 control-flow instructions
 are taken during the execution of the whole program. In these cases, the overhead is below 1.8.
On the other hand, for \aesb, the overhead increases by a factor of up to 4.6, due to the very high number of control-flow instructions
taken (over 506). 
Overall, in our benchmarks, executing secured functions incurs costs of up to a factor of 4.6 compared to the original, non-secured code.
Of course, these numbers need to be weighed against the frequency at which the secured function would be used in a real-life scenario. }

\begin{figure}[t]
  \includegraphics[scale=1.4]{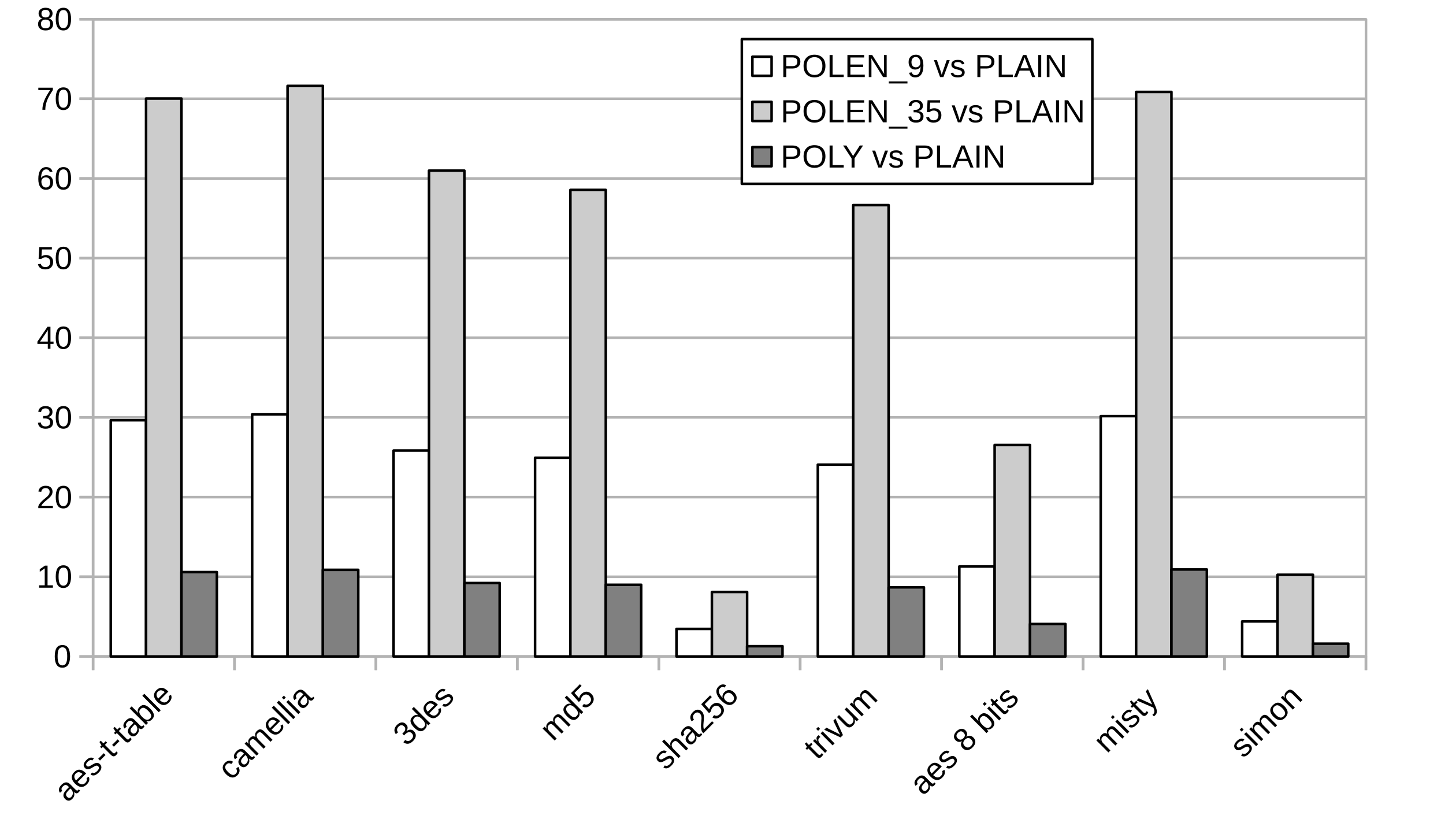}
  \caption{Execution time overhead for the generation of polymorphic instances, for \polyC, \polenC\_9 and \polenC\_35, expressed relative to the execution of the unprotected function in \plainC{} mode.}
  \label{fig:cost of generation poly polen vs plain}
\end{figure}

\dmr{
Second, we evaluate the cost of generating a new polymorphic instance compared to executing an unprotected version of the original function. Results are shown in Figure~\ref{fig:cost of generation poly polen vs plain}. 
Overheads are reported for \polyC, \polenC\_9{} and \polenC\_35 (from white to light and dark grey, respectively) compared to \plainC{}. 
\sha, \aesb{} and \simon{} provide the lowest overheads of 8.1, 26.55 and 10.25, while other cases peak at a factor of 70 (for \polen\_35). The latter finding is due to the presence of highly iterative loops. In practice, the cost of generating a loop is of the order of magnitude of the number of instructions contained in the loop body.
On the other hand, the cost of executing the generated loop depends both on the number of instructions contained in the loop body \textit{and} on its iteration bounds. As an example, the \sha{} encrypt function contains two loops totalling 16 operations, which need to be generated dynamically (one operation corresponds to multiple machine instructions), in addition to the loop control structure.
However the corresponding loop bounds lead to the execution of the same operation several hundreds of times, which makes runtime code generation much cheaper w.r.t.\ execution time.
}

\begin{figure}[t]
  \centering
  \subfloat[Overhead $\mathcal{O}$ measured on polymorphic instances.  $\mathcal{O}$ varies depending on the code of the secured function, in particular, its control structure.\label{fig:overhead instances}]{
    \includegraphics[width=.45\textwidth,keepaspectratio,valign=t]{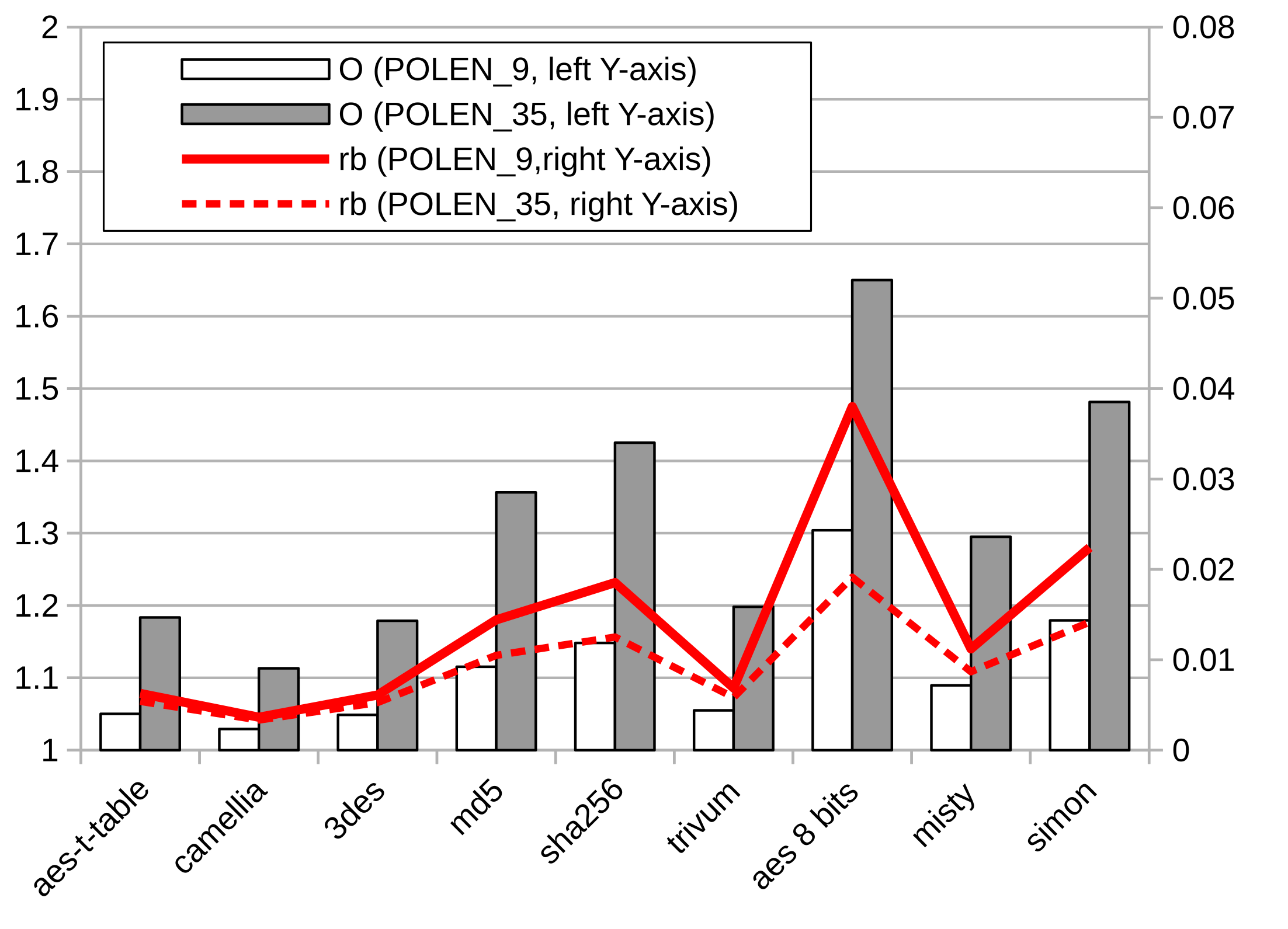}
      }
  \qquad
  \subfloat[Overhead  $\mathcal{O}$ measured on SGPCs.
  Here, $\mathcal{O}$ is stable across all cases as the control structure of the SGPC is very regular from one case to another.\label{fig:overhead generators}]{
    \includegraphics[width=.45\textwidth,keepaspectratio,valign=t]{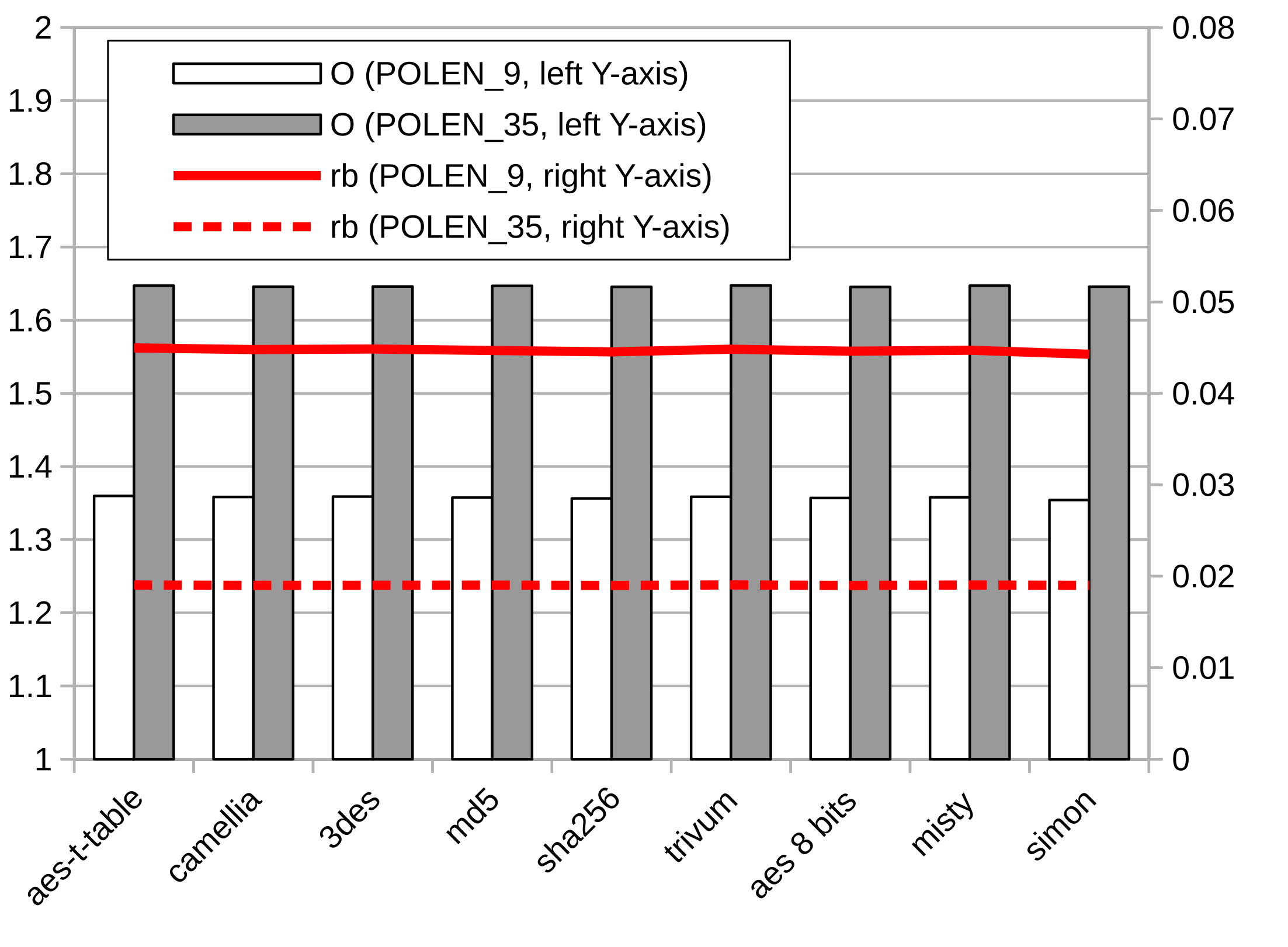}
  }
  \caption{Overhead due to code encryption, for \polenC\_9 and \polenC\_35.
Bars show the overhead $\mathcal{O}$ (left Y-axis) as a percentage of the cost of executing the non-encrypted version.
The $rb$ ratio is shown as a plain or dotted line (right-Y-axis).
}
  \label{fig:overhead encrypt}
\end{figure}

\dmr{Finally, we evaluate the performance overhead due to encryption alone.
Results are shown in Figure~\ref{fig:overhead encrypt}. 
In the following, $n$ denotes the total number of instructions executed for the secured function; $b$ is the number of 
control-flow instructions taken; 
$i = n - b$ is the number of other instructions; 
and $\kt{}$ denotes the cost of initialisation of the Trivium modules, expressed in CPU cycles
(as a reminder, in our simulation model, each executed instruction takes one CPU cycle).
In our evaluation, $\kt{}$  can be either 9 (\polenC{}\_9) or 35 (\polenC{}\_35).
We measure the overhead due to encryption as 
$\mathcal{O} = \frac{E_\mathcal{E}}{E_{\phi}}$, where $E_\mathcal{E}$ denotes the execution time with decryption activated, and $E_{\phi}$ denotes the execution time without decryption activated.
This overhead is due to the use of code encryption, but in the following we illustrate that it is impacted by the control-flow structure of the secured program, and that it impacts the use of code polymorphism as well.
$\mathcal{O}$ can be also modelled as $\mathcal{O} = \frac{i+ \kt{} \times b}{i+b} = 1 + (\kt{}-1) \times rb$, 
where $rb$ denotes the ratio $\frac{b}{n}$ or the number of control-flow instructions taken per instruction executed.
That is, with code decryption, each taken control-flow instruction incurs an execution time penalty of $\kt{}$ due to the reinitialisation of the stream cipher.
%
%
For polymorphic instances (Figure~\ref{fig:overhead instances}),
variations of the measured $\mathcal{O}$  follow the variations of $rb$, which supports the validity of our model.
In our benchmarks,  $\mathcal{O}$ reaches a maximum of 1.65 for \aesb{} with $\kt{}=35$
due to the very high number of control-flow instructions taken, similarly to our results in Figure~\ref{fig:overhead execution time of polen instances}.
The measure of $\mathcal{O}$ for SGPCs (Figure~\ref{fig:overhead generators}) shows a consistent overhead from one program to another.  
This suggests that the control-flow structure of SGPCs is rather independent of the nature of the target program to secure.
This also implies that the major decision factor for SGPCs is the cost of the initialisation of the Trivium modules.  
We further discuss this point in Section~\ref{sec:Duplicating Trivium Instances}.
}

In general, we observe that the overhead is mainly influenced by: 
i)~$\kt{}$, the cost of initialisation of the Trivium module; 
ii)~the number of basic blocks in the secured code; 
iii)~the average size of a basic block; 
and iv)~the presence of loops in the code (in particular, when these loops are highly iterative).
Those observations show that, for simple processors, the overhead is mainly influenced by the control flow graph and not the type of instructions.


%

\subsubsection{Hardware Overhead}
\label{ssec: hw criteria}

The \polen{} architecture was evaluated on a functional simulator for the RV32IM ISA, designed as an extension to the \spike{} instruction set simulator.
A drawback of this decision is that we were not able to assess the exact hardware overhead in terms of logic gates.
However, we have tried to provide estimates of costs, based on results published in the literature.
Previous work on an Intel Cyclone-V FPGA~\cite{Hiscock2019} reported that a 32-bit Trivium represents 21\% (237 Adaptative Logic Modules (ALMs), which are configurable logic blocks in Intel FPGAs) of a 32-bit MIPS CPU, a processor with a 5-stage integer pipeline.
For the same processor, a 128-bit Trivium represents 67\% (1,094 ALMs) of the CPU surface.
Thus, if \polen{} was implemented on the same processor, the hardware overhead would be around 21\% for an \emph{area-optimised} version and 67\% for a \emph{performance-optimised} version.
That being said, we expect that a real application would occupy more area than the toy processor used in the literature~\cite{Hiscock2019}.
For example the VexRiscv~\cite{vexriscv} CPU, a 32-bit RISC-V core, which is capable of running Linux, requires 1,764 ALMs on a Cyclone-V FPGA.
The relative Trivium (and, thus, \polen{}) overheads on this target would be of the order of 13\% for a 32-bit Trivium and 41\% for a 128-bit Trivium.

\section{Discussion}
\label{sec:discussion}

\subsection{Duplicating Trivium Instances}
\label{sec:Duplicating Trivium Instances}
Conceptually, \polen{} requires two Trivium instances: one for decrypting incoming instructions, and another to encrypt polymorphic instances.
Thus, the designer has to choose between having \dmr{two distinct hardware} instances or a single one that is shared between the instruction fetch and the execution stages. For in-order scalar processors, two instances do not seem to be interesting as not only are hardware costs significant relative to the core, but we do not expect huge benefits in terms of performance. Moreover, a scalar processor would be unable to execute instructions while waiting for encryption to complete. However, an out-of-order processor may benefit from having a separate instance, as the core would be able to fully overlap encryption with the execution of other instructions.

Furthermore, the evaluation of the overhead $\mathcal{O}$ in the preceding section shows that it is constant for SGPCs, because their structure (in particular the control flow graph) is independent of the structure of the target polymorphic program. This supports different design decisions for the two Trivium instances.
For example, it is possible to integrate a very efficient Trivium module for decryption, e.g., with a higher unrolling factor and $\kt{} \ll 9$,
which would reduce $\mathcal{O}$ at the expense of a larger silicon footprint.
This can also be compensated for by the use of a less efficient encryption module (e.g., $\kt{}=35$), which is only used for encryption in the SGPCs:
%
it should be noted that in many cases, SGPCs can be executed far less often than polymorphic instances, meaning that the extra cost of the latter can easily be compensated for.

\subsection{Software or Hardware Trivium Instance}

The previous section discussed whether or not a distinct hardware Trivium instance should be used for encryption. However, another possibility is to implement the encryption module as a software library.
\dmr{This solution doesn't require an ISA extension, meaning that the \polen{} portability is improved and that the overall hardware cost is reduced.}
Of course, in this case, encryption execution time would be far longer compared to a hardware primitive. The main drawback is that encryption keys are managed in software, and may reside in memory. Therefore, this countermeasure offers weaker protection.
As an illustration, Dropbox client versions 1.1 to 1.2.8 performed software encryption and opcode permutation in the Python bytecode. Nevertheless, \citeauthor{kholia2013looking} managed to bypass the encryption and recover the opcode mapping~\cite{kholia2013looking}.

%
%
%




\subsection{Code Encryption with a Block Cipher}
\label{sec:block ciphers}
The implementation of \polen{} presented in Section~\ref{sec:evaluation} uses stream \dmr{ciphers.
In some contexts, it may be required to use \polen{} with a block cipher (e.g., the developers already have an hardened implementation).
The mapping of common cipher modes to the abstract cipher model defined in Section~\ref{sec:encryption_formalism} is straightforward.} 
For example, to use a 128-bit AES in counter mode, the $IV$ and $state$ would both be the counter value (a 128-bit value). Then, \polen{} would be instantiated with:
\begin{itemize}
    \item $\mathcal{I}_k(IV) = IV$
    \item $\mathcal{T}_{enc, \: k}(state, m) = (state + 1, AES_k(state) \oplus m)$
    \item $\mathcal{T}_{dec, \: k}(state, m) = (state + 1, AES_k(state) \oplus m)$, is the same as $\mathcal{T}_{enc}$
\end{itemize}

Nevertheless, switching to a block cipher has several important implications for \polen{}:
\begin{itemize}
    \item A block cipher operates on blocks that are much larger than instructions (e.g., 128 bits) and
        additional buffering hardware will be required to manage this.  
        This limitation also implies the use of padding when the size of basic blocks does not fit with 128-bit boundaries.
    \item The operations $\mathcal{T}_{enc, \: k}$ and $\mathcal{T}_{dec, \: k}$ are likely to have much higher latency and decrypting at the instruction granularity of the CPU might not be possible. It could also be the case that latency is so high that it is not possible to decrypt one or more instructions per clock cycle.
    \item The operation $\mathcal{I}_k(IV)$ may be significantly faster than with a stream cipher, which means that basic block merging heuristics presented in Section~\ref{ssubsec:cfg prep} will have to be changed.
      \dmr{Ultimately, if $\mathcal{I}_k(IV)$ incurs a negligible overhead, basic block merging can be disabled, because control-flow instructions have no extra penalty.}
\end{itemize}
\dmr{Those reasons highlight that stream-ciphers are a sound default choice for \polen{}.}

\subsection{Encryption Scope}
\label{sec:discussion enc scope}
The ability to move in and out of the encryption domain is of particular interest when the protected call uses shared or system libraries. However, protecting such functions raises many questions.
First, ensuring that all of the shared functions in the encrypted code can themselves be encrypted requires access to the source code in order to compile an encrypted version of the binary. Second, aside from their use in protected code, functions in shared libraries may be called by unprotected code. In this case, both encrypted \textit{and} unencrypted versions of these functions would be required; however, keeping both versions is a security breach in itself as it provides plaintext/ciphertext combinations that an attacker can use to guess the encryption key. Finally, protecting too many functions in the application codebase may severely impair its overall performance, as can be anticipated from the results presented in Section~\ref{subs:performance}. During deployment, the programmer can configure \polen{} to the specific usage setting.

\section{Related Works}
\label{sec:related works}

\polen{} protects both code and code pointers through encryption.
Although sensitive data is not encrypted (only programs are encrypted), it is protected from a SCA by code polymorphism, which implements a form of side-channel hiding protection.
%
%
While this combination protects software against each attack vector, more importantly, it also protects against more complex forms of attacks that benefit from each vector.
Compared to architectures with memory encryption, such as AEGIS~\cite{Suh2005} or Intel SGX~\cite{costan2016intel}, \polen{} mitigates both side-channel and code-extraction threats and responds to calls in recent studies that advocate for comparable combinations of hardware and software countermeasures.
In Polyglot~\cite{Sinha2017}, Instruction Set Randomisation is combined with code encryption, and is shown to increase resistance to code reuse attacks, including Just-In-Time Return-Oriented Programming. The binary of the protected application is diversified before deployment to a particular device. It is then encrypted offline with one encryption key per memory page and decryption is triggered by the operating system, on page loading, within the system's MMU. 
We believe that \polen{} and Polyglot have comparable security properties concerning code reuse attacks, although \polen{} has not been validated specifically against this type of attack. In Polyglot, decryption is performed at the frontier between the CPU and the cache hierarchy. As far as we know, no attack has been demonstrated that exploits a data leak from the MMU's internal processor. Therefore, Polyglot's approach seems sufficient to address code confidentiality. 
\dmr{Polyglot also seems to be easier to integrate into existing processors than \polen{}, which requires modifications to the CPU core. An interesting line of work would be to adapt code polymorphism to memory-level encryption with Polyglot, rather than our current, in-core decryption. 
In this case, extra care would be needed to ensure the confidentiality of the datapath that is used to write \polen's dynamically generated instructions to memory. }

\dmr{More recently, Morpheus~\cite{Gallagher2019} combines encryption of code, code pointers, and data pointers with the creation of two separate, randomly displaced, address spaces for code and data above the virtual address space.
The encryption countermeasure is inserted between the L1-L2 cache boundary, and cache tags are used to select the associated encryption keys.
Similarly to \polen{}, Morpheus supports code encryption, but the demonstrated overhead is much lower (around 1\% penalty with re-randomisation periods of 50~ms). 
We believe that this is due to the fact that, in Morpheus, the encryption latency is masked by the latency of L2 cache accesses, while in \polen{} we do not assume any memory architecture and the decryption module is located in the processor micro-architecture, which also protects against attackers capable of observing the contents of the L1 memory caches.
Another major difference is that the Morpheus architecture is supported by \emph{domain tagging}: each execution \emph{domain} is associated with 2-bit tags, which are associated to each program instructions.  The whole processor micro-architecture is modified to propagate tags to the output values of each instruction and to operate with the \emph{pointer displacement} and \emph{domain encryption} defences.
In contrast, \polen{} only requires the addition of a decryption and an encryption modules to the fetch and decode stages of the processor pipeline, respectively.
Last, Morpheus exploits the virtual memory mechanism to introduce some form of code randomisation of the code and data memory layouts.
In \polen{}, code polymorphism is implemented in software, which incurs a higher overhead, but supports a wider range of randomisation capabilities, at the level of machine instructions.}

\dmr{
It has also been observed that software obfuscation is vulnerable to timing SCA, mainly because the fine-grained timing of programs is generally predictable~\cite{Fell2019}.
Fell et al.\ remove the conditional control-flow instructions vulnerable to timing attacks, and replace some of the original program instructions with so-called \textit{custom instructions}.
Custom instructions are supported by dedicated hardware which introduces at runtime non-deterministic variability in the execution time.
Fell et al.'s use of custom instructions is also one of the goals of code polymorphism: to introduce runtime variability to mitigate side-channel attacks.  
However, our implementation of code polymorphism does not require dedicated hardware support.
In \polen{}, code polymorphism protects against side-channel attacks, 
and code encryption protects against reverse engineering.
Code encryption is stronger than any form of code obfuscation as long as the attacker cannot access to the encryption key.
Please note that, in our work, code polymorphism is used as a countermeasure against side-channel attacks, and not as an obfuscation countermeasure.
The runtime variability provided by code polymorphism could be used as a form of mitigation against reverse engineering, but this would require further work to properly assess the benefits of code polymorphism as a form of code obfuscation.
}

\section{Conclusion}
\label{sec:conclusion}

In this paper, we consider that an attack scenario involving side-channel analysis is the
combination of \dmr{two successive phases}:
an analysis phase and 
an exploitation phase.
We advocate that practical protections against such attacks need to
address these two phases.
As a countermeasure against the analysis phase, our approach considers the use of code encryption.
As a countermeasure against the exploitation phase, our approach considers the use of code \dmr{polymorphism. This technique} relies on runtime code generation, which makes its combination with code encryption particularly challenging.

We have presented a combination of code encryption with code polymorphism, as implemented in the \polen{} toolchain.
Code encryption protects against all forms of attacks that rely on reverse engineering the static, binary version of the attacked program.
Our framework can encrypt all code produced statically (including runtime code generators) or dynamically (polymorphic instances) for target functions.
Therefore, our approach significantly strengthens programs against advanced attacks that rely on both reverse engineering and side-channel information extraction.

We measured the performance of \polen{} against a number of representative cryptographic implementations.
Security was evaluated using simulated side-channel traces for an AES software implementation protected with \polen{}.
Our leakage analysis showed that our countermeasures made it more difficult to identify points of interest in side-channel observations, and correlation analyses were noticeably more difficult with the addition of code polymorphism.
In particular, our security evaluation illustrates that encryption alone does not protect against side-channel attacks, and emphasises the importance of combining it with other protections.

From the programmer's point of view, \polen{} is easy to use as it only requires the developer to define which parts of the code (either statically or dynamically generated) should be encrypted, and what sources of variability should be used to generate polymorphic variants.
The configurability of our approach makes it possible to fine-tune the application of both countermeasures, and weigh security/performance trade-offs, depending on the application's needs and context.

\begin{acks}
This work was supported by the French program ``Programme Investissements d'Avenir IRT Nanoelec'' \grantnum{ANR}{ANR-10-AIRT-05},
and the European project SERENE-IoT (\grantnum{PENTA}{16\,004})  within the framework of PENTA, the EUREKA cluster for Application and Technology Research in Europe on NanoElectronics.
\end{acks}

\bibliographystyle{ACM-Reference-Format}
\bibliography{polen,Damien}

\end{document}